\documentclass[aps,pra,floatfix,superscriptaddress]{revtex4}

\UseRawInputEncoding

\usepackage{mathrsfs}
\usepackage{graphicx}
\usepackage[english]{babel}
\usepackage{amsmath}
\usepackage{amssymb}
\usepackage[colorlinks,linkcolor=blue]{hyperref}

\usepackage{relsize}
\usepackage{CJK}
\usepackage{dsfont}
\usepackage{bm}

\newcommand{\me}{\mathrm{e}}

\newcommand{\dif}{\mathrm{d}}

\allowdisplaybreaks
\begin{document}

\title{Machine learning of XY model on a spherical Fibonacci lattice}
\author{Chen-Hui Song }
\affiliation{Department of Physics, Southeast University, Jiulonghu Campus, Nanjing 211189, China}
\affiliation{Tsinghua Shenzhen International Graduate School, Tsinghua University, Shenzhen 518055, China}
\thanks{Southeast university and Tsinghua university provide equal support to the majority of this work.}
\author{Qu-Cheng Gao}
\affiliation{Department of Physics, Southeast University, Jiulonghu Campus, Nanjing 211189, China}
\author{Xu-Yang Hou}
\affiliation{Department of Physics, Southeast University, Jiulonghu Campus, Nanjing 211189, China}
\author{Xin Wang}
\affiliation{Department of Physics, Southeast University, Jiulonghu Campus, Nanjing 211189, China}
\author{Zheng Zhou}
\affiliation{Department of Physics, Southeast University, Jiulonghu Campus, Nanjing 211189, China}
\author{Yan He}
\affiliation{School of physics, Sichuan University, Chengdu, Sichuan 610064, China}
\author{Hao Guo }
\email{guohao.ph@seu.edu.cn}
\affiliation{Department of Physics, Southeast University, Jiulonghu Campus, Nanjing 211189, China}
\author{Chih-Chun Chien}
\affiliation{Department of physics, University of California, Merced, CA 95343, USA}

\begin{abstract}
We study the XY model on a spherical surface inspired by recently realized spherically confined atomic gases. Instead of a traditional latitude-longitude lattice, we introduce a much more homogeneous spherical lattice, the Fibonacci lattice, and use classical Monte Carlo simulations to determine spin configurations.
The results clearly show that topological defects, in
the form of vortices, must exist in the stable configuration on a sphere but vanish in a plane due to a mathematical theorem. 
Using these spin configurations as training samples, we propose a graph-convolutional-network based method to recognize different phases, and successfully predict the phase transition temperature. We also apply the density-based spatial clustering of applications with noise, a powerful machine learning algorithm, to monitor the merging path of two vortices with different topological charges on the sphere during Monte Carlo simulations. Our results provide reliable predictions for future space-based experiments on ultracold atomic gases confined on spherical lattice in the microgravity environment.

\end{abstract}

\maketitle

\section{Introduction}

The XY model\cite{XY61}, although very simple, is of great importance to a lot of quantum many-body systems, such as the liquid helium, antiferromagnetic insulators, and superconductors\cite{Betts77}. In the two-dimensional (2D) situation, the XY model has inspired the research boom 
on topological defects and unconventional phase
transitions\cite{KT73,KT74,KT16}. During the last half century, there have been tons of studies on various aspects of XY model\cite{XY70,XY74,XYPRB79,XYPRB92,XYOlsson92,XYOlsson95a,XYOlsson95b,Nagaosa,Altland,WenXG,XYJSM05,XYJCP09,XYPRE19,XYPRR21}. Experimentally, the XY model can be emulated by ultracold atoms confined in optical lattices\cite{XYNP13}. Recently,
the fast development of space-based technique has
stimulated experimental efforts to confine ultracold
atoms on a spherical surface in microgravity\cite{SphericalBECNJP20,SphericalSFPRL20,SphericalSF21,SphericalBECPRL19,SphericalBECnpj19,SphericalBECPRA21,SphericalBECS10,SphericalBECN18}, making a systematic analysis of the XY
models on a spherical lattice an important task. However, 
in contrast to
the planar case, there is no arbitrarily large and exactly uniform lattice on a sphere, which prevents a direct analytical study on such models.
Suitable numerical methods must be developed to make reliable and instructive predictions. Fortunately, this can be achieved by the aid of powerful tools in processing big data, such as the machine learning. During the last several years, this method has been successfully introduced to the physics community from computer sicent, and achieved tremendous and unexpected progresses\cite{MLPRB16,MLS17,MLNP17,MLPRX17,MLPRX17b,MLPRB17b,MLPRB17c,MLPRB17d,MLPRE17,MLPRE17b,MLPRE18,MLPRL18a,MLPRL18b,CNNPRApp20,CNNJAP20}.

In this paper, we focus on the spherical XY model. Instead of the traditional latitude-longitude lattice, we first introduce an approximately uniform lattice, the Fibonacci lattice, on the surface of a sphere, which is basically the most uniform lattice on a spherical surface. Then, with the help of Monte Carlo (MC) simulations, we find the spin configurations of these spherical XY models.
Multiple vortices onset even at ultralow temperatures and the net topological charge is always 2 due to the topology of 2D sphere $S^2$.
Since the spherical Fibonacci lattice still lacks translational invariance, the traditional convolutional neutral network (CNN) can not be directly applied here, we instead introduce the graph-convolutional-network (GCN) method. Using the samples of spin configuration provided by MC simulations, GCN successfully predict the Berezinskii-Kosterlitz-Thouless (BKT) phase transition temperature. In some sense, the MC annealing of a quantum system is equivalent to the spontaneous relaxation dynamics, during which the vortices may move and merge with each other.
We then apply the density-based spatial clustering of applications with noise (DBSCAN) to give the dynamical trajectory of these vortices, which shows the merging of vortices with different topological charges during MC annealings. Our results and methods provide a solid theoretical reference for future space-based experiments on spherically confined ultracold atomic gases.

\section{The Spherical XY model}
\subsection{The XY Model}

The 2D classical XY model on a square lattice is described by the Hamiltonian
 \begin{align}\label{H}
H=-J\sum_{<i,j>}\mathbf{s}_i\cdot\mathbf{s}_j=-J\sum_{<i,j>}\cos(\theta_i-\theta_j)
\end{align}
where $J$ is the interaction constant, $\mathbf{s}_i$ denotes the spin with angle $\theta_i$ at site $i$, and the sum is taken over all pairs of nearest-neighbor spins. At the zero temperature, the system stays at the ground state in which all spins are aligned in the same direction. There are also excitations with nontrivial topology, such as vortices and anti-vortices, which onset below the
BKT transition temperature $T_\text{c}$.
An interesting question is: What happens when all spins are confined on topologically nontrivial surfaces like $S^2$?

\begin{figure}[th]
\centering
\includegraphics[width=2.8in]{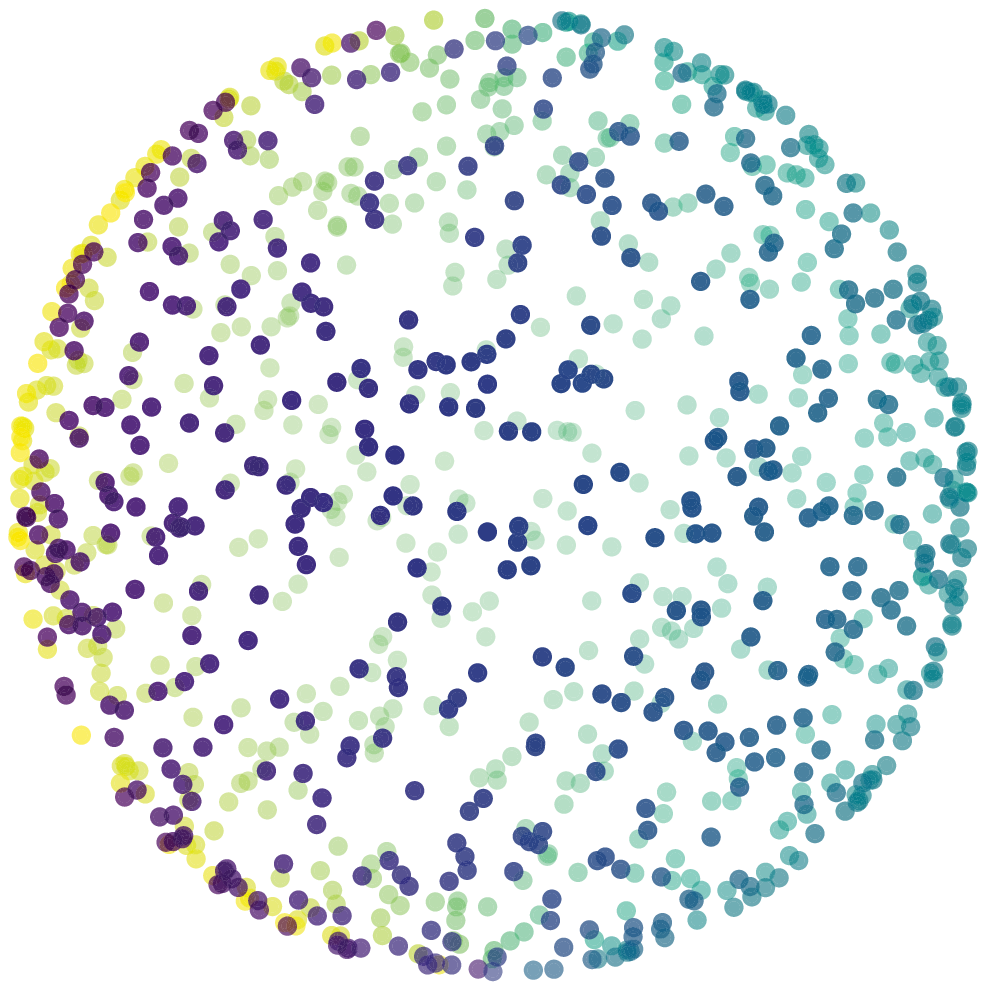}
\includegraphics[width=2.8in]{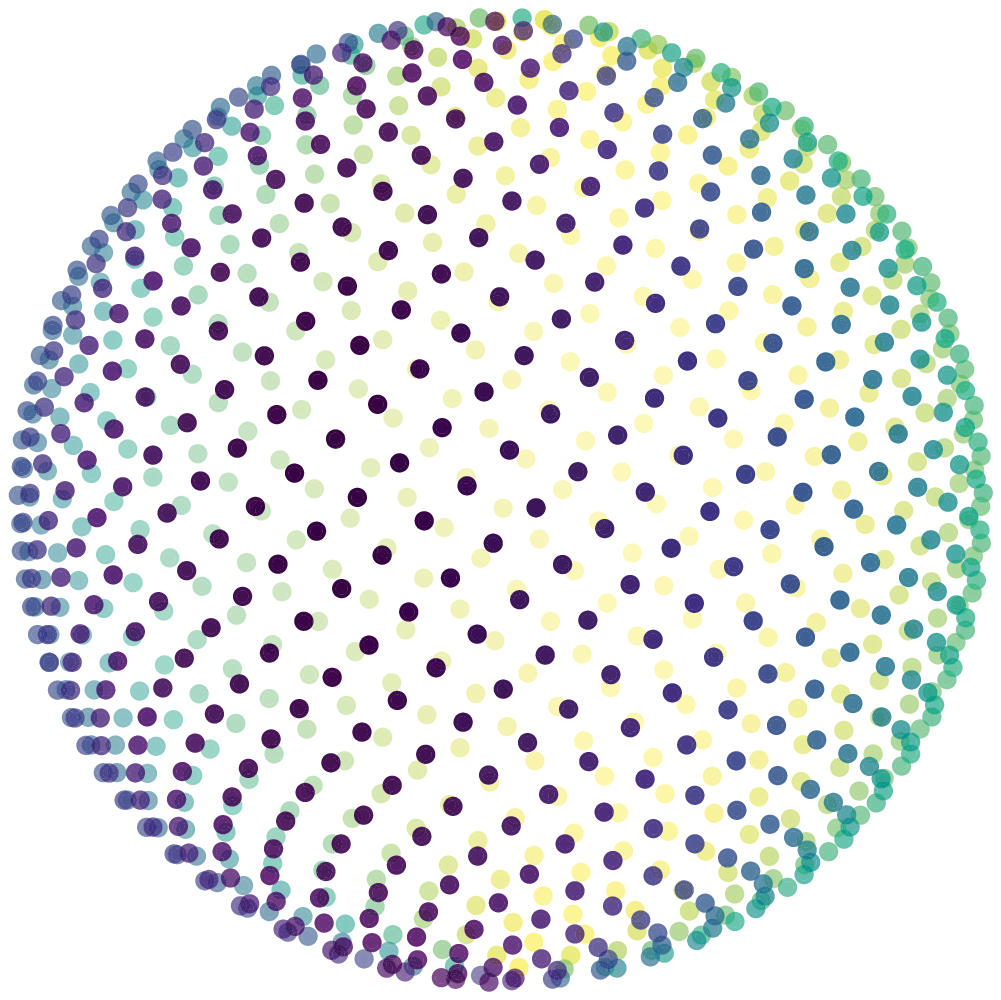}
\caption{(Left panel) Random lattice on a spherical surface. (Right panel) Fibonacci lattice on a spherical surface.}
\label{Fig1}
\end{figure}

\subsection{The Fibonacci Lattice}

Note the map of a XY model onto a sphere is nontrivial since the 2D plane is topologically different from a spherical surface. On a square lattice, the ordinary XY model described by Eq.(\ref{H}) is isotropic and the lattice site is evenly distributed. When confined on a spherical surface, some important issues must be carefully treated. For example, how can we arrange the spins as homogeneous as possible?
How is the Hamiltonian revised with respect to a certain spherical lattice? Obviously, the expression on the far right-hand-side of Eq.(\ref{H}) does not apply any more since the translational invariance is not preserved here.
A regular and convenient choice of coordinate lattice seems to be the latitude-longitude lattice. However, it is highly inhomogeneous: The site density near the north or south pole is much denser than anywhere else. If a spherical lattice
is totally isotropic, its sites must be located at the vertices of a regular polyhedron. There are only five different types of regular polyhedrons, among which the regular dodecahedron has the largest number of vertices: 20. In other words, a totally isotropic spherical lattice can only have at most 20 sites, which is still too small for a systematic study of
the XY model towards the thermodynamic limit.
For
a lattice with a large number of sites, we have to seek
an alternative allowing the area occupied by each site
to be almost identical. 
Fortunately, there exists such a spherical lattice with a large number of sites, called the Fibonacci lattice~\cite{Fibonacci,FibonacciMG}, where the $i$-th site on a sphere of radius $R$ is defined by
 \begin{align}\label{FL}
x_i=\sqrt{R^2-z^2_i}\cos(2i\pi\phi),y_i=\sqrt{R^2-z^2_i}\sin(2i\pi\phi), z_i=R\left(\frac{2i-1}{N}-1\right),
\end{align}
for $i=1,2,\cdots, N$.
Here $N$ is the total number of sites, and $\phi=\frac{\sqrt{5}-1}{2}$ is the golden ratio. When $N$ is large, the $i=1$, $N$ sites approach the southern and northern poles of the sphere respectively. To visualize the difference between the random and Fibonacci latttices on a sphere, we present in Figure \ref{Fig1} a comparison between the two types of lattices, both with $N=1000$ sites. Apparently, the site distribution of the latter is much more uniform. Basically, the Fibonacci lattice is the most uniform among all lattices on a sphere. 

Since the spherical Fibonacci lattice is not strictly isotropic, we introduce a cutoff distance $r_\text{c}$ such that the interaction is allowed only when the separation between each pair of spins is less than $r_\text{c}$. Moreover, the distance between an arbitrary site and its nearest neighbour is not a constant, and we accordingly assume a Gaussian type interaction
 \begin{align}\label{HS}
H=-J\sum_{<i,j>}\me^{-\alpha r^2_{ij}}\mathbf{s}_i\cdot\mathbf{s}_j
\end{align}
where the constant $\alpha$ determines the scale of the short-range interaction, and $r_{ij}$ is the distance between the sites $i$ and $j$. Each spin is normalized to be unit-lengthed, and is locally confined on the tangent plane at its site.
To demonstrate the uniformity of Fibonacci lattice, we consider a unit sphere sprinkled with a Fibonacci lattice of $N=1000$ sites and set $r_\text{c}/R=0.11395$. For each site, the lattice points that fall in its domain of radius $r_\text{c}$ are referred to its neighbours. It is found that 850 sites have 4 neighbours, 76 sites sites have 3 neighbours, and 74 sites have 5 neighbours. Altogether, there are $850\times 4+76\times 3+74\times 5=3998$ neighbours, and interactions are only allowed between these 3998 pairs of spins. Basically, this is almost the best approximation to a 2D square lattice for spherical lattices, where each site has exactly 4 nearest neighbors. The distribution of pair separations are outlined in Table.\ref{T1}.
\begin{table}[th]
  \centering
\begin{tabular}{|c|c|c|c|c|c|}
  \hline
  \text{Number of pairs} & \text{Average distance} & \text{Standard deviation} & \text{Median} & \text{Minimum} & \text{Maximum} \\
  3998 & 1.1455 & 0.0816 & 1.1194 & 0.9777 & 1.2977 \\
  \hline
\end{tabular}
  \caption{Distribution of pair separations.}\label{T1}
\end{table}

\subsection{Effect of the topology of $S^2$}
\begin{figure}[th]
\centering
\includegraphics[width=2.5in]{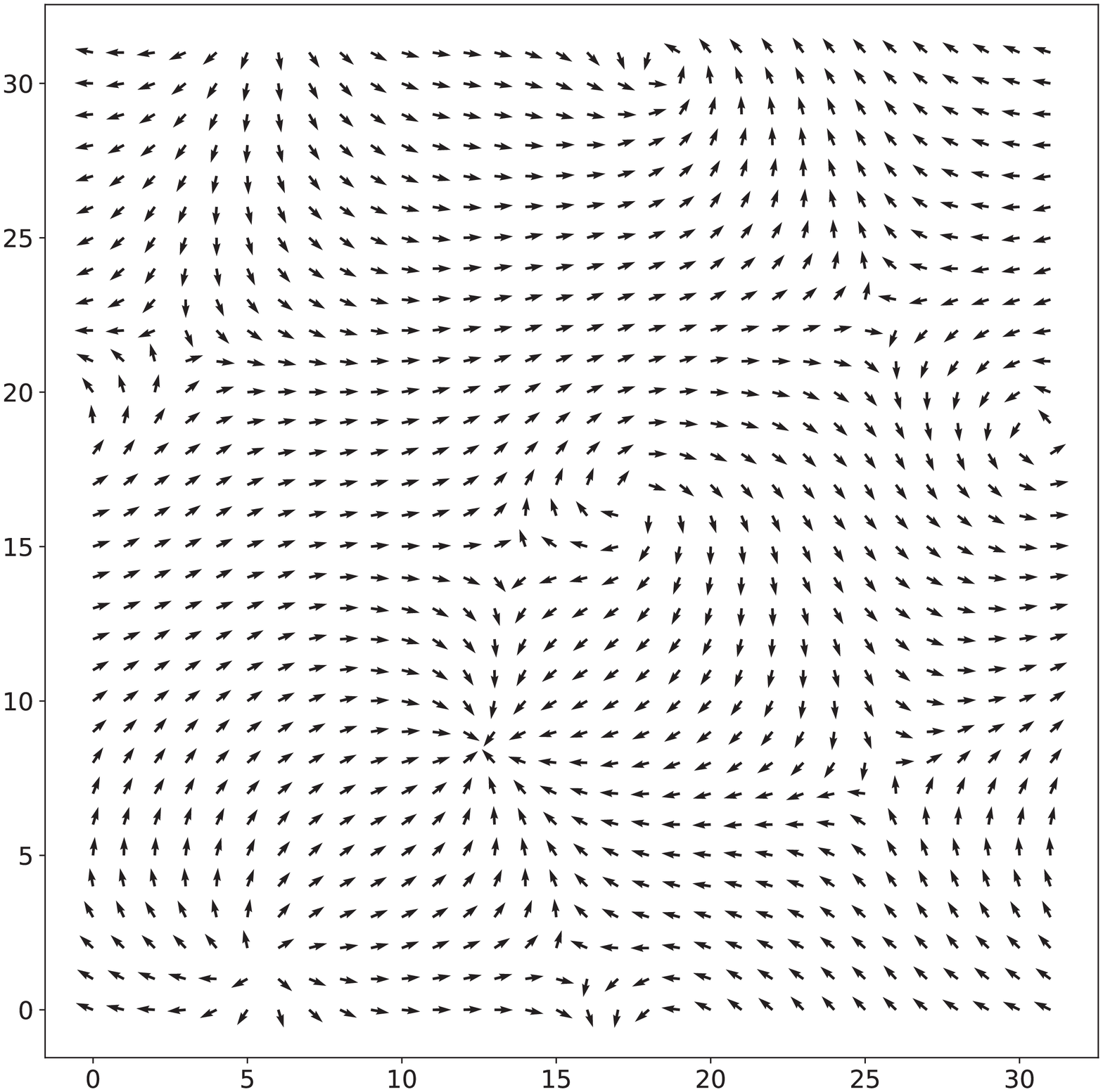}
\includegraphics[width=2.5in]{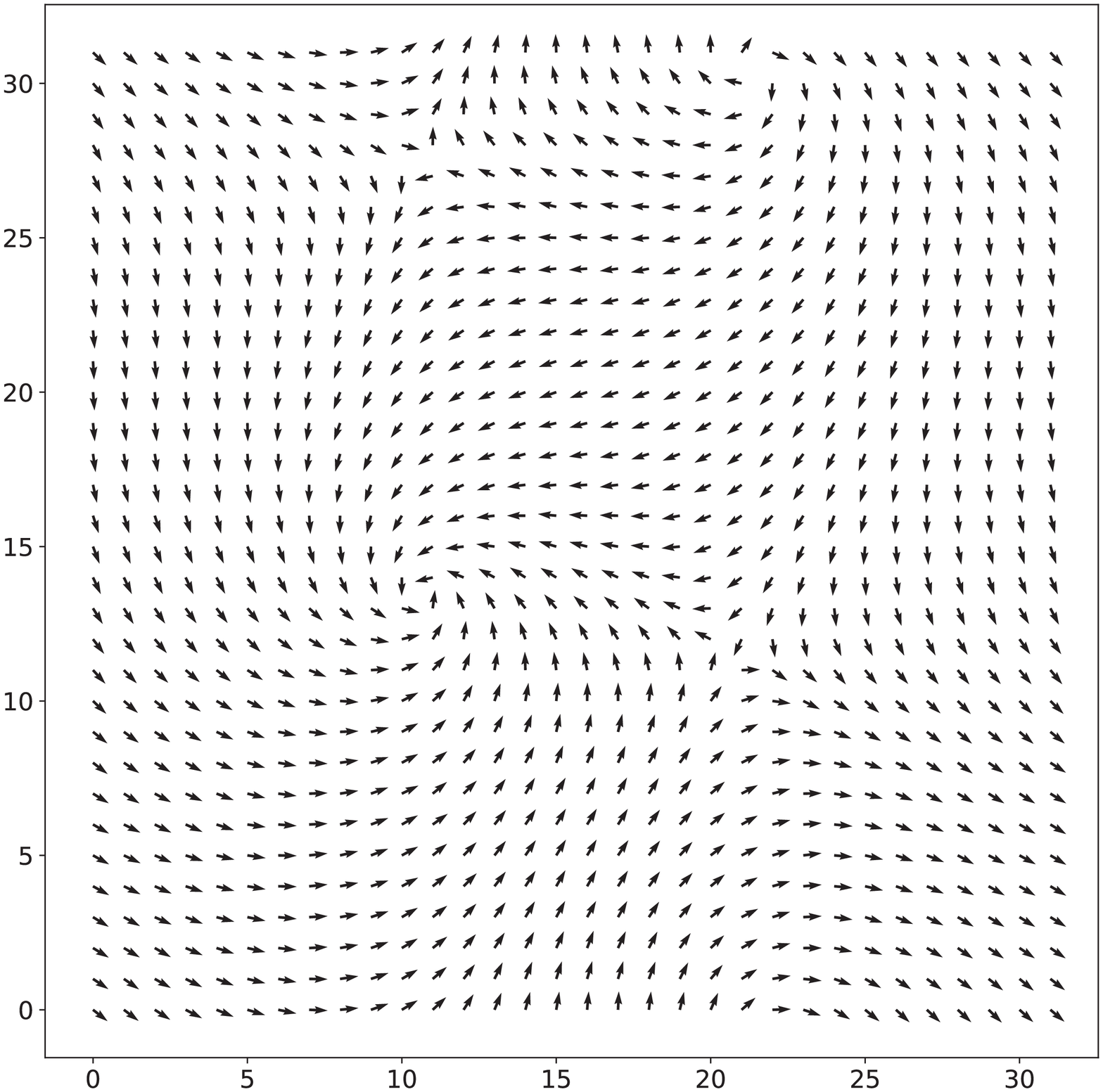}
\caption{Snapshots of the spin configurations of the XY model on a $32\times 32$ 2D square lattice from Monte Carlo simulations with $r_\text{c}/a=1.0$ (left), 
 and $r_\text{c}/a=2.0$ (right). Here $T/J=5.0\times 10^{-3}$. The vortex density decreases as the interaction range increases.}
\label{Fig2}
\end{figure}

We first give an overview of some key points of the XY model on a 2D square-lattice.
At zero temperature, the ground state is a topologically trivial state with $\theta_i=$constant. Away from the zero-temperature limit, topological defects called the ``vortices'' appear in the form of bound  pairs with opposite topological charges.
As temperature increases, the vortex-antivortex pairs start to unbind at the Kosterlitz¨CThouless (KT) temperature $T_\text{c}$~\cite{KT73}. To compare with the spherical-lattice case, we also include a Gaussian-type interaction to the XY model on a square lattice and introduce $r_\text{c}$ to control the range of interaction. We implement MC simulations to study the XY model on a 32$\times$32 square lattice with lattice constant $a$. The vector field of the spins are visualized in Fig. \ref{Fig2}. In the left panel, we set $r_\text{c}/a=1.0$ and the model reduces to an ordinary XY model. It can be seen that several vortices appear in bound pairs. In the right panel, $r_\text{c}/a=2.0$. We find that the number of vortices decreases significantly. This is because $r_\text{c}$ determines the scale of local order. As it increases, the pair size increases due to the longer range of correlation. Thus, the visible vortex pairs per area reduce correspondingly.

When mapped onto a spherical surface, since all spins are confined on the tangent plane at each site, the spins actually belong to a tangent vector field $X$ on $S^2$, which is nondegenerate. The Poincar$\acute{\text{e}}$-Hopf theorem\cite{PHT} states that
 \begin{align}\label{PH}
\sum_i\text{ind}_{x_i}(X)=\chi(S^2).
\end{align}
Here $x_i\in S^2$ denotes the zeros of the vector field $X$, $\text{ind}_{x_i}(X)$ means the index of $X$ at $x_i$, and $\chi(S^2)$ is the Euler characteristic of $S^2$, which is 2. The zero $x_i$ in fact corresponds to a vortex, since the center of a vortex is a zero point of a vector field. Moreover the index at $x_i$ equals to the product of its sign and multiplicity, where the sign corresponds to the topological charge, and the multiplicity is usually 1. Physically, the Poincar$\acute{\text{e}}$-Hopf theorem indicates that the net topological charge of a spherical-lattice XY model is always 2. This result is independent of temperatures, hence vortices must onset even at zero temperature, i.e. in the ground state, which is essentially different from the square-lattice XY model.

\begin{figure}[th]
\centering
\includegraphics[width=2.2in]{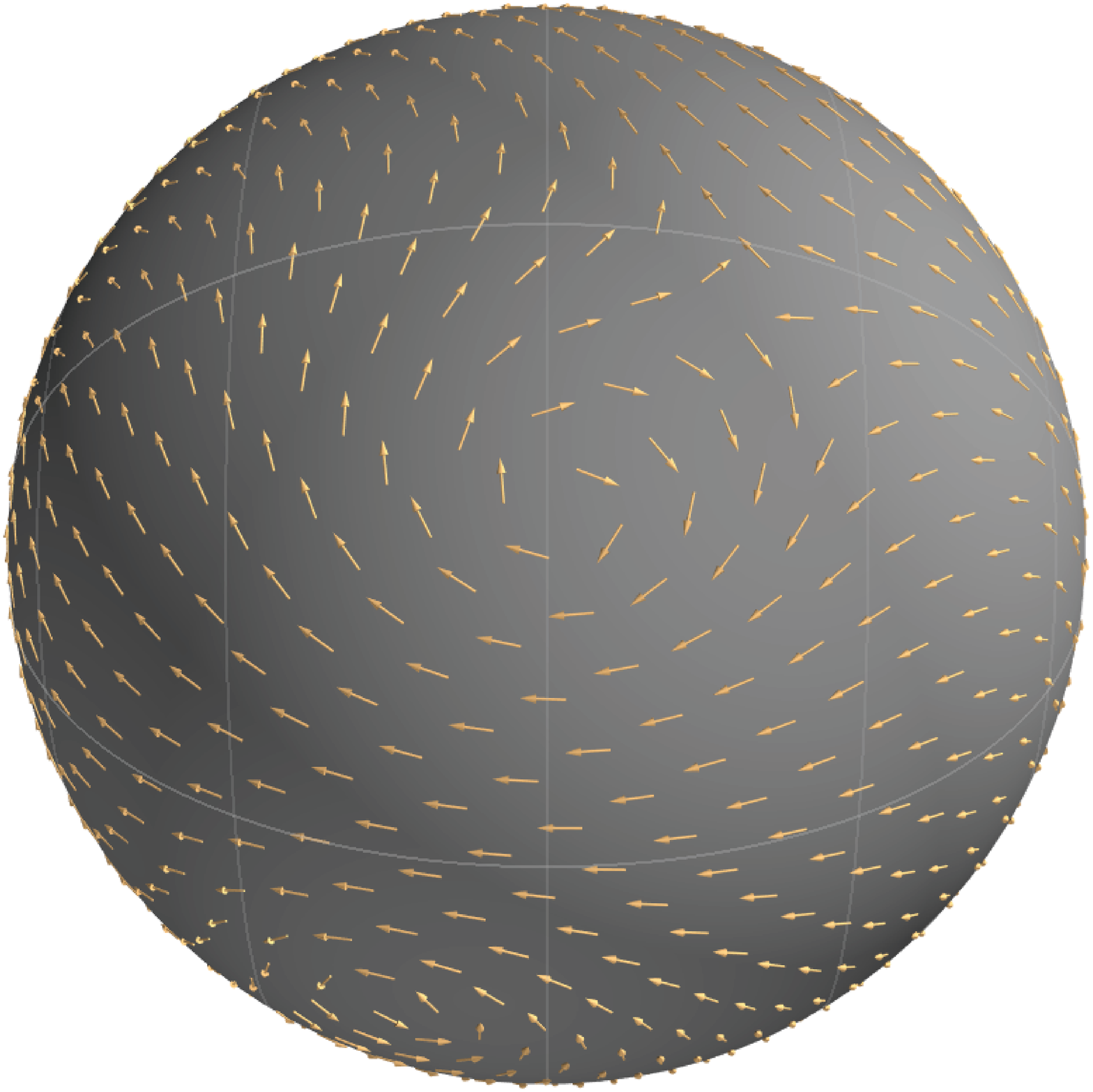}
\includegraphics[width=2.2in]{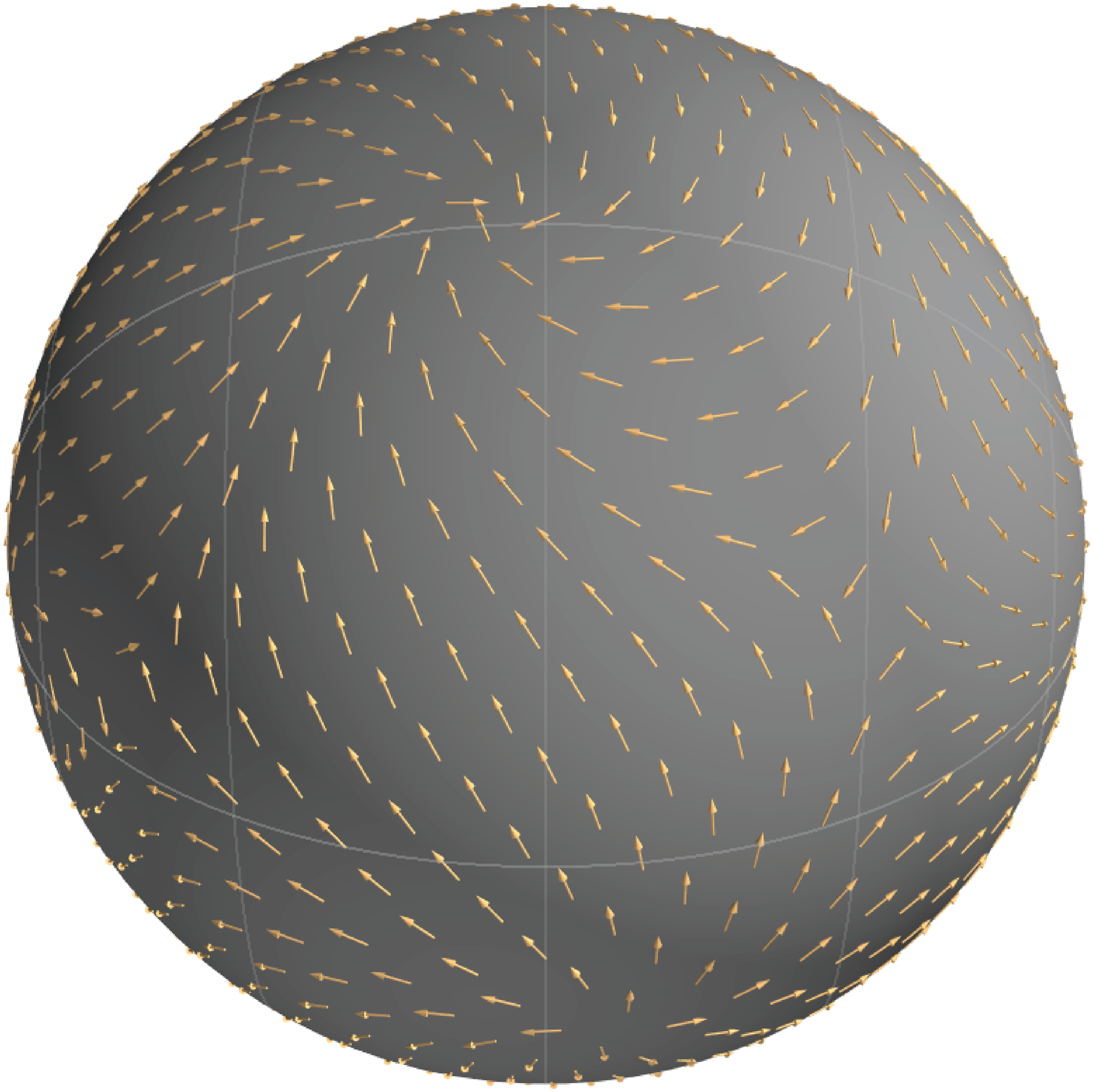}
\caption{Monte Carlo simulations of the XY model on a spherical Fibonacci-lattice after a partial annealing of $5\times 10^6$ steps. Here $N=1000$, $T/J=5.0\times 10^{-4}$, and $r_\text{c}/R=0.11395$. The left (right) plot shows the front (rear) side, chosen arbitrarily.}
\label{Fig3}
\end{figure}

\begin{figure}[th]
\centering
\includegraphics[width=1.9in]{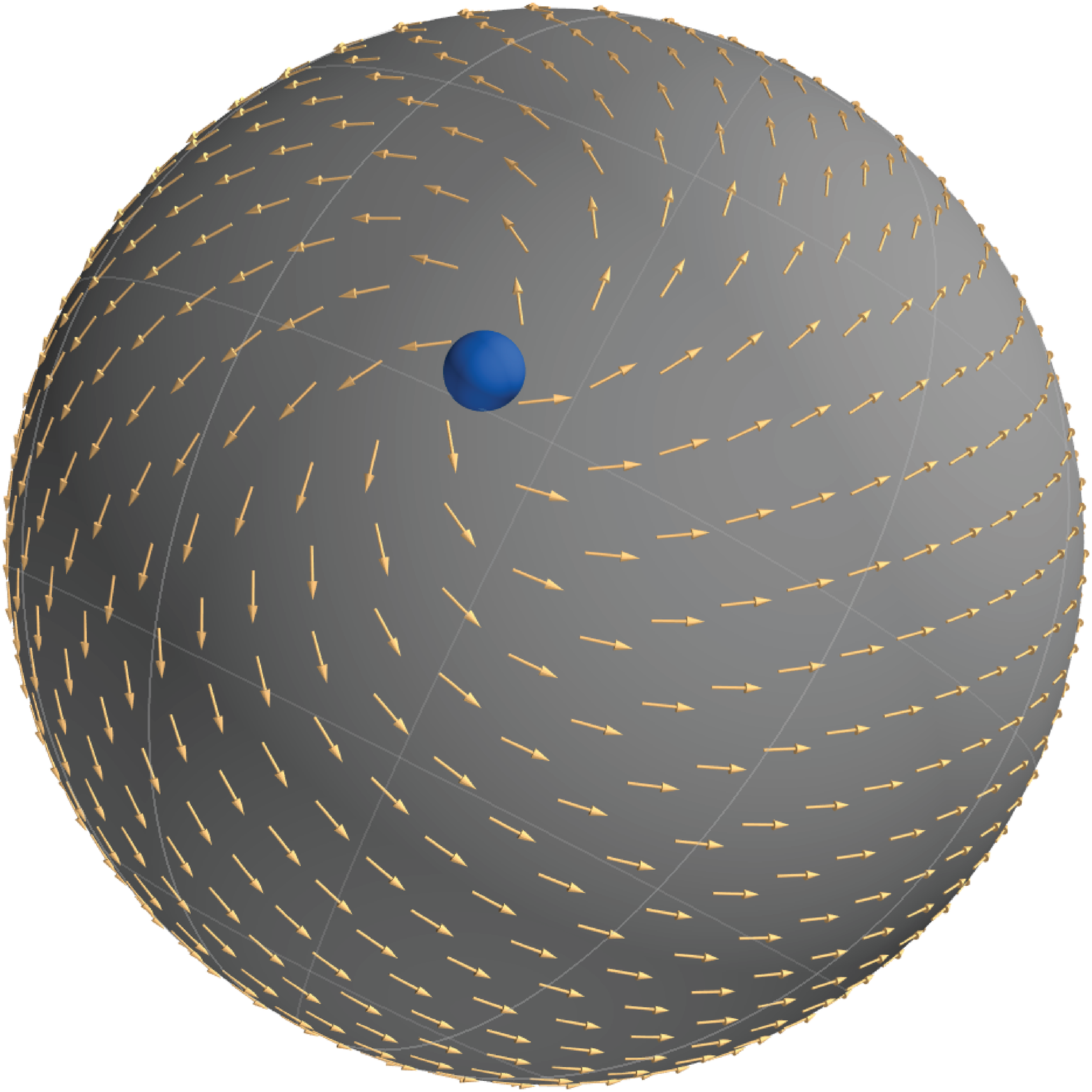}
\includegraphics[width=1.9in]{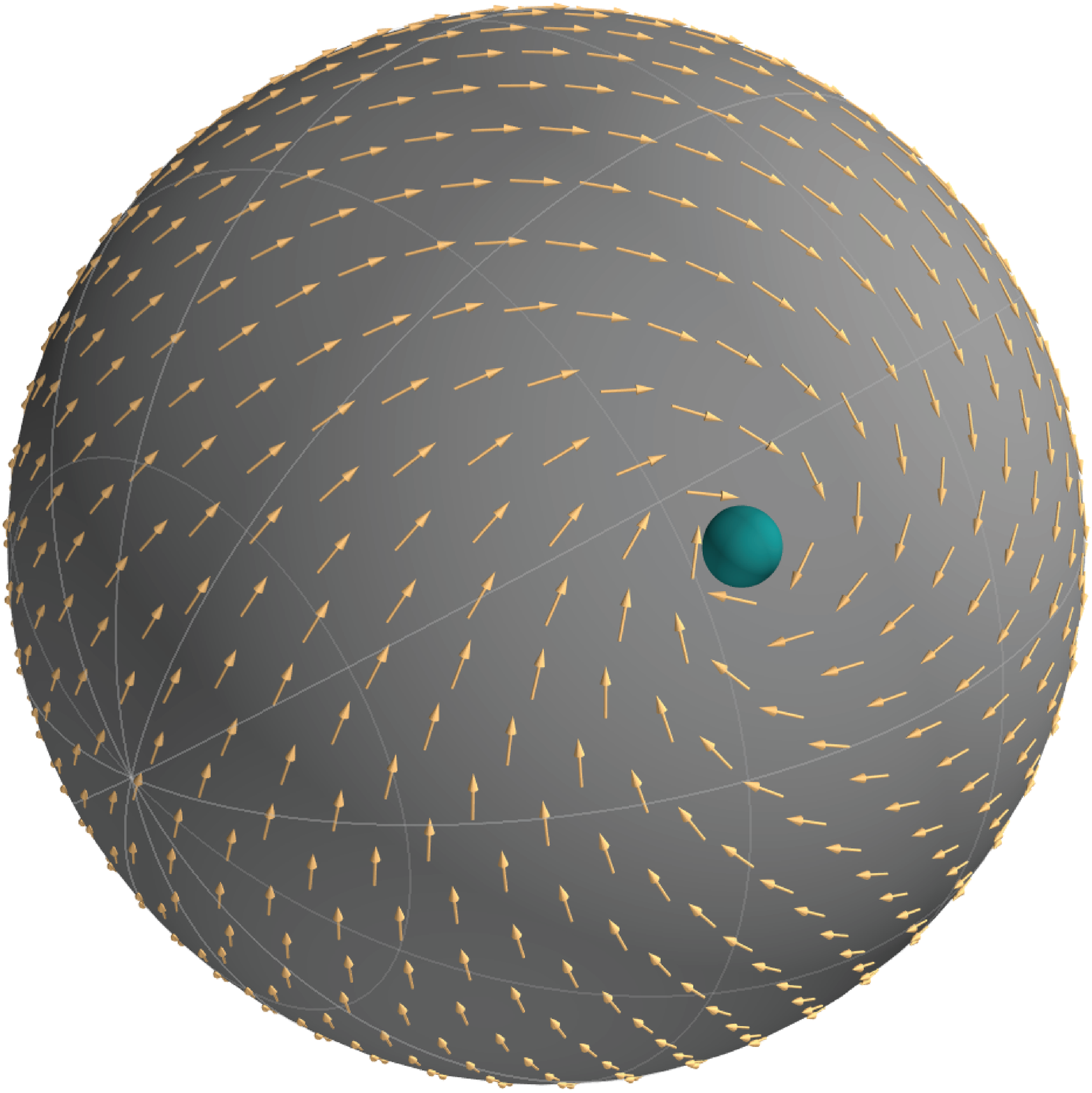}
\includegraphics[width=1.9in]{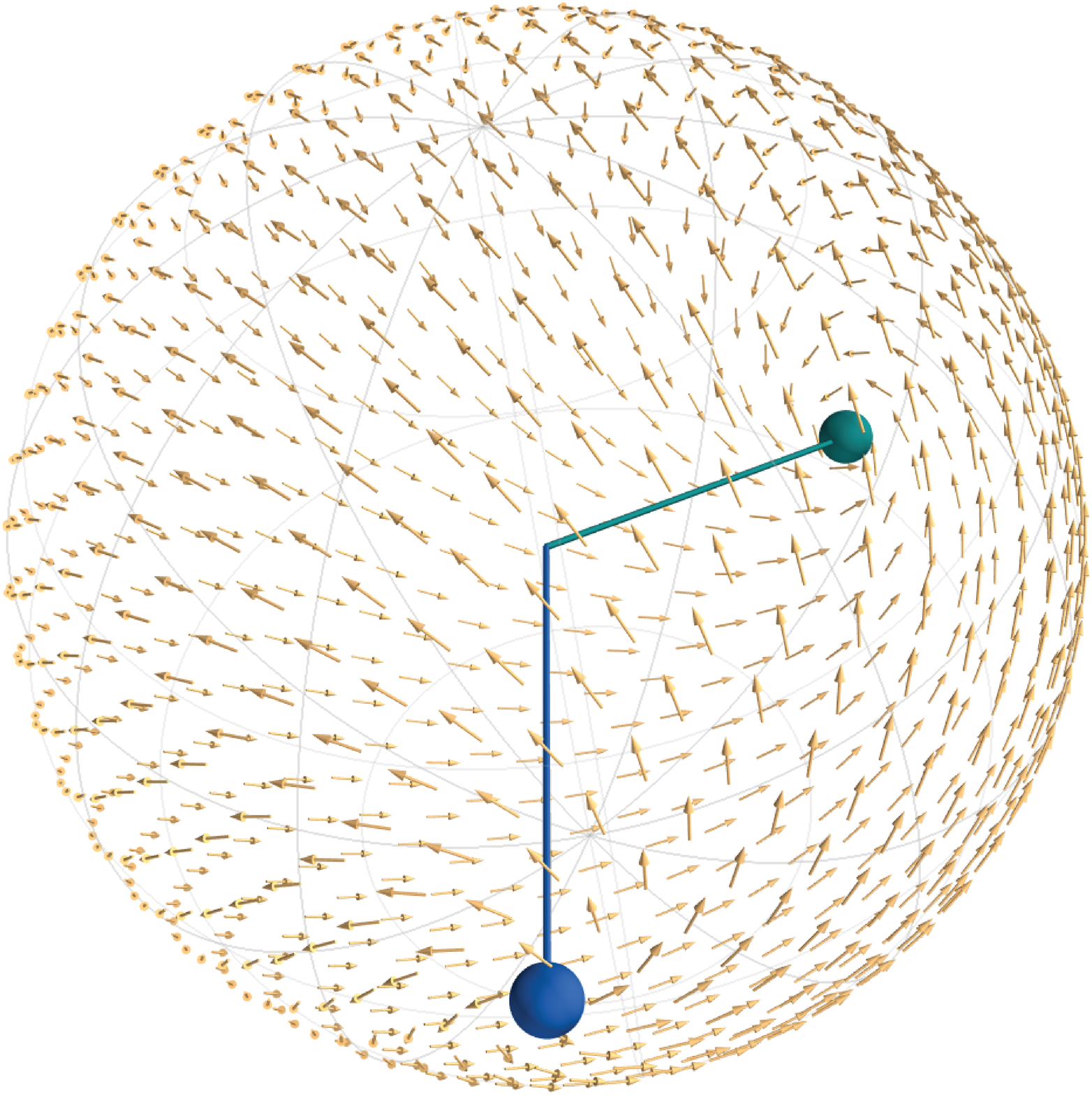}\\
\includegraphics[width=1.9in]{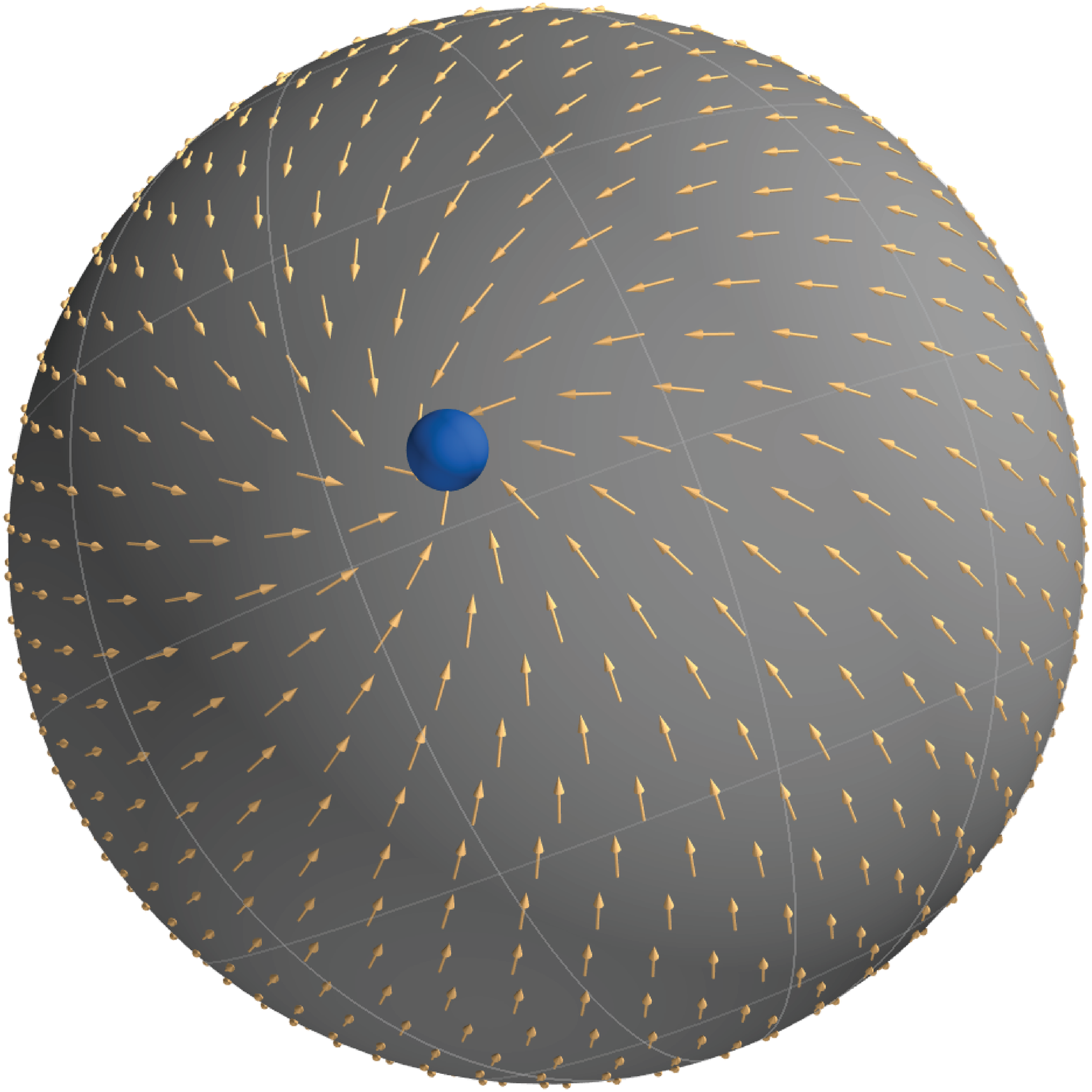}
\includegraphics[width=1.9in]{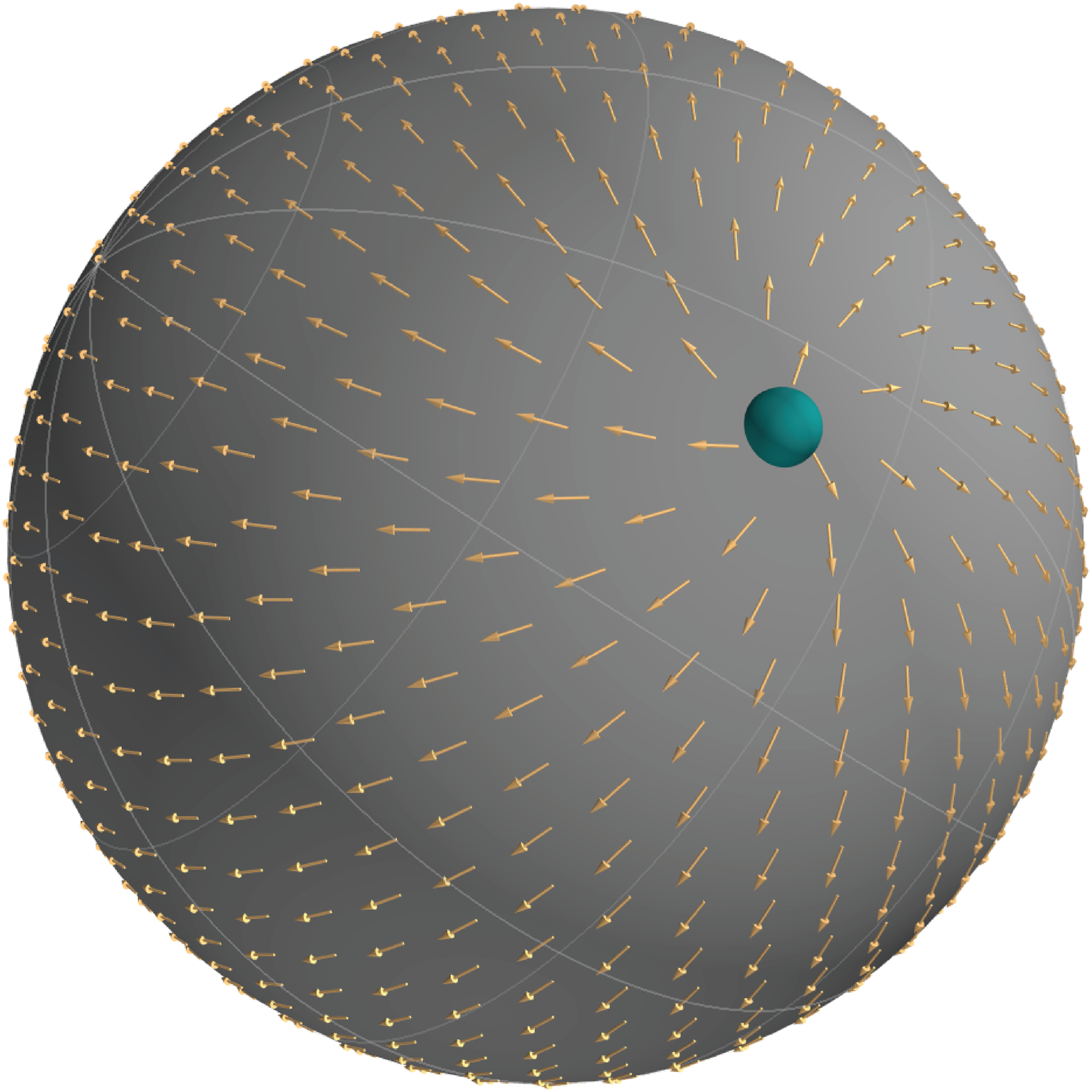}
\includegraphics[width=1.9in]{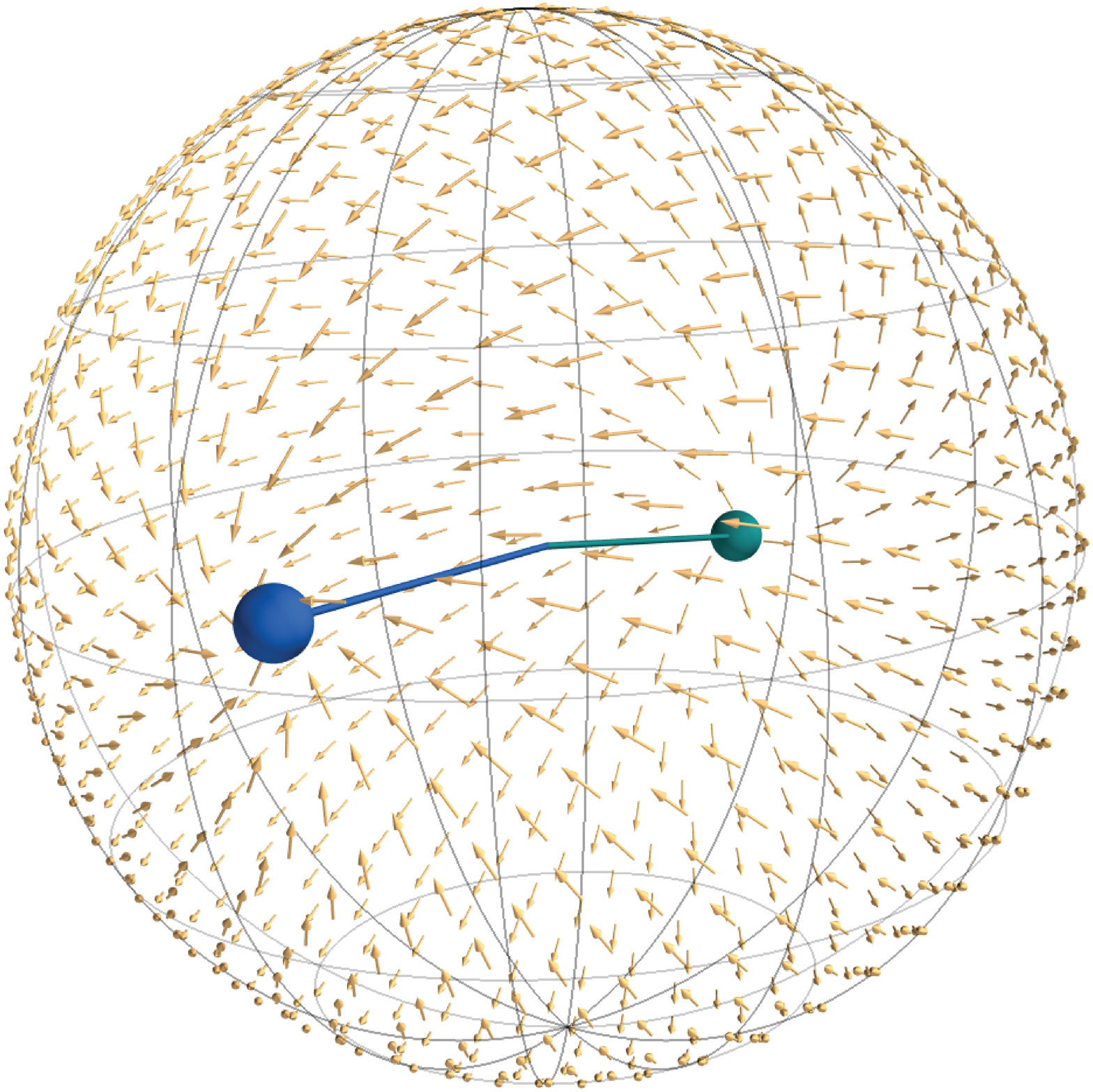}
\caption{Monte Carlo simulations of the  XY model on a spherical Fibonacci lattice after a complete annealing. Top row: (Left) A vortex on one side. (Middle) Another vortex on the opposite side.  (Right) Perspective drawing. Here $r_\text{c}/R=0.11395$. The angle between the two vortices with respect to the origin is $131.5^{\circ}$. Lower row: (Left) A vortex on one side. (Middle) Another vortex on the opposite side. (Right) Perspective drawing. Here $r_\text{c}/R=0.2$. The angle between the two vortices is $174.3^{\circ}$. }
\label{Fig4}
\end{figure}

\section{Spin Configurations}

\begin{figure}[th]
\centering
\includegraphics[width=1.9in]{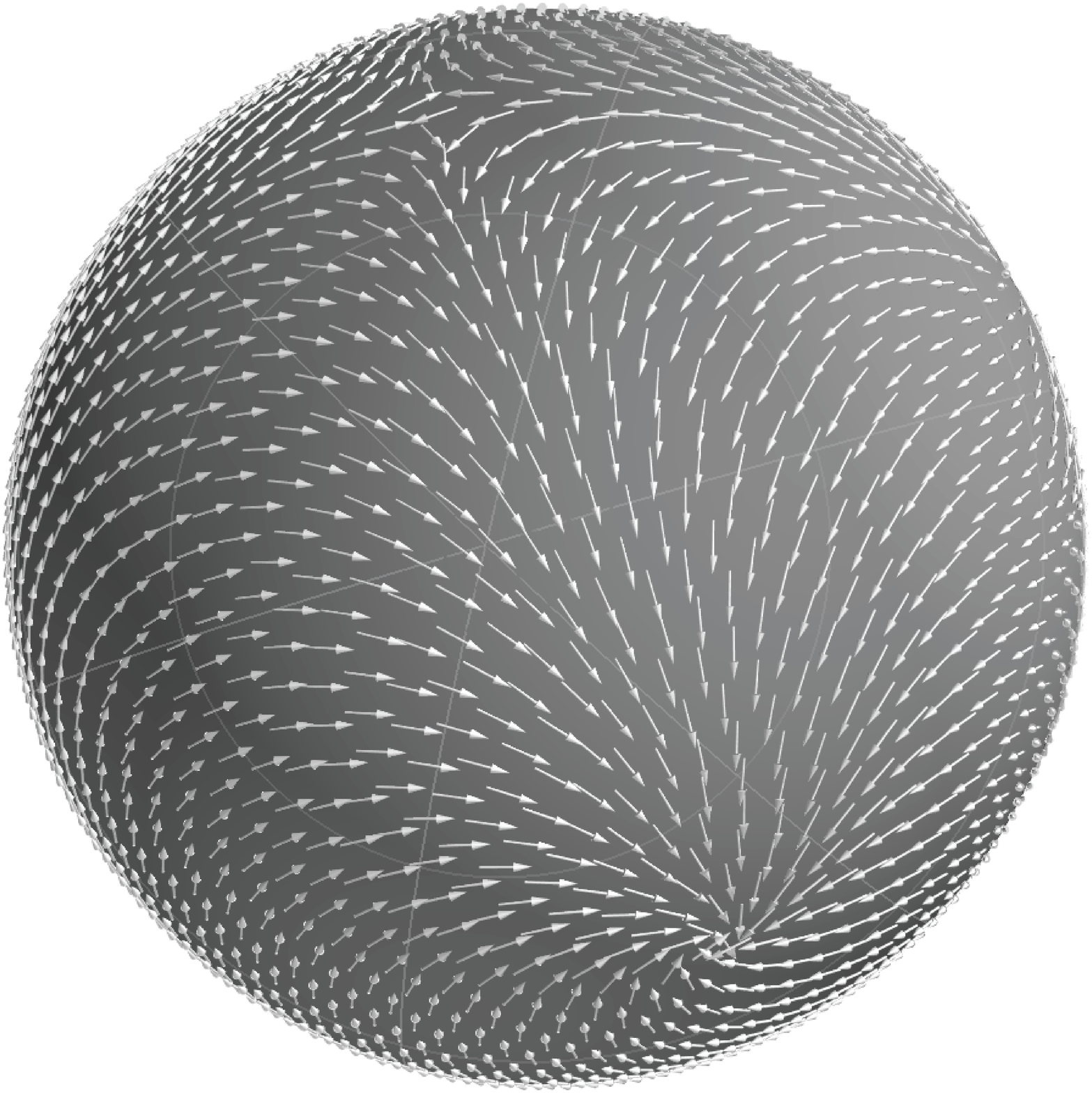}
\includegraphics[width=1.9in]{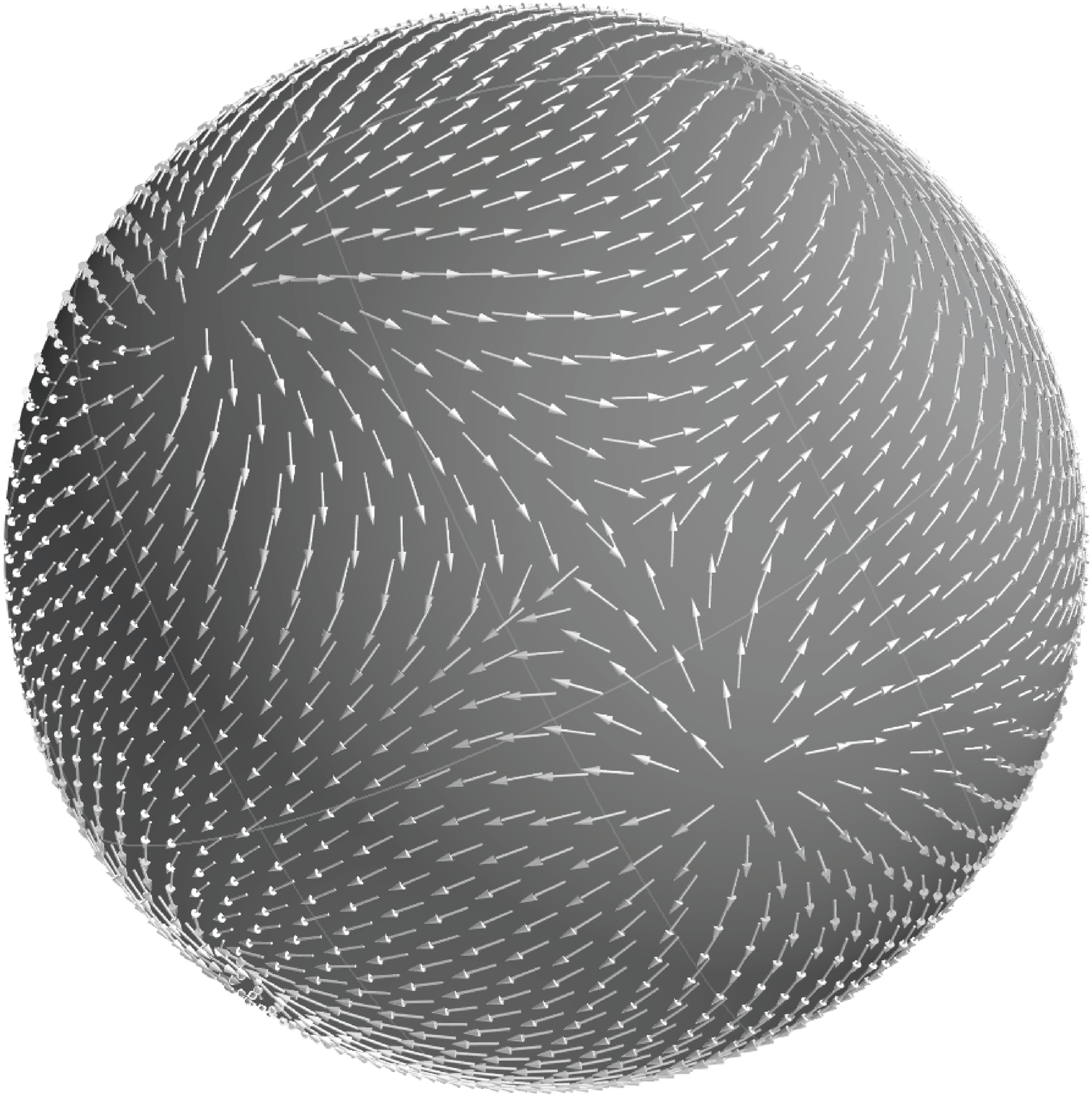}
\includegraphics[width=1.9in]{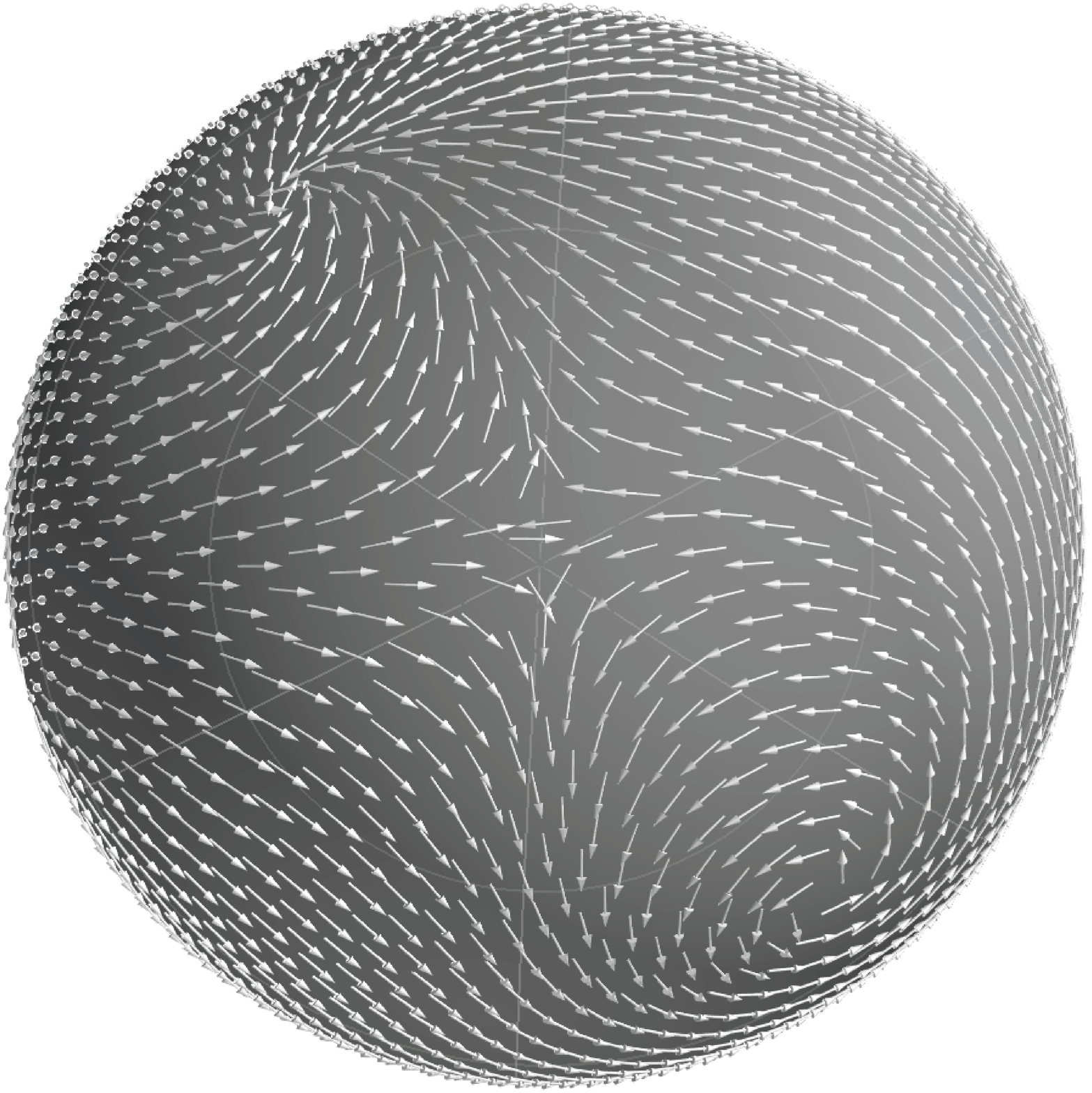}
\caption{Monte Carlo simulations of the XY model on a spherical Fibonacci lattice  after a complete annealing. Here $N=3000$, $T/J=5.0\times 10^{-4}$, and $r_\text{c}/R=0.09$. The spin configuration is presented from three different directions to show all vortices. }
\label{Fig4b}
\end{figure}

We then consider the XY model on a Fibonacci lattice on a spherical surface of radius $R$. By choosing $R$ as the length unit, we consider the unit sphere in our work. The cases with $N=1000$, $\alpha=10/R^2$, $T/J=5.0\times 10^{-4}$, and $r_\text{c}/R=0.11395$ or 0.2 will be presented. The spins are randomly oriented initially and then evolve under the MC annealing afterwards. The simulation runs 1.5$\times 10^8$ steps until the spin configuration is stable. It can be found that multiple vortices appear during the annealing. Figure \ref{Fig3} presents an intermediate state when the annealing runs $5\times 10^6$ steps, and there are a few vortices. As the simulation continues, some vortices merge with one another. We show the stable spin configuration in Fig. \ref{Fig4} after running 1.5$\times 10^8$ steps and find that only two unpaired vortices survive. Moreover, these two vortices tend to move away from each other during the annealing, and the angle between the two survived vortices with respect to the origin is $131.5^{\circ}$ for $r_\text{c}/R=0.11395$. If $r_\text{c}/R=0.2$, the two vortices finally reside at two nearly opposite sites on the sphere with the angle between the two vortices given by $174.3^{\circ}$, but not necessarily at the two poles. This is shown in the perspective drawings in Fig. \ref{Fig4}. The topological charge of a vortex is given by
 \begin{align}\label{TC}
n=\frac{1}{2\pi}\oint \nabla\theta\cdot \dif\mathbf{l},
\end{align}
where the integral is evaluated along a closed curve encircling the center of the vortex. Using this formula, it can be found that both vortices have the same charge $+1$, which agrees with our previous analysis based on the Poincar$\acute{\text{e}}$-Hopf theorem. This also explains why the vortices move as far as possible from each other: It is because the repulsion between them. The number of vortices of the final stable state actually depends on the choice of the parameters. In Figure \ref{Fig4b}, we show the final stable spin configuration by choosing $N=3000$, $T/J=5.0\times 10^{-4}$, and $r_\text{c}/R=0.09$ after the annealing runs $1.5\times 10^8$ steps. In this case, there are eight vortices, five of which have charge +1, three have charge -1 (For more details, please refer to the video mentioned below). Thus, the net topological charge is also +2, just as expected. This is quite reasonable. In fact, the $N=3000$ lattice becomes much larger than the $N=1000$ case with respect to a typical pair of adjacent sites. Thus, the distance between the vortices is effectively larger and the interaction becomes relatively smaller, which may prevent the merging of vortices when the annealing goes on. We also made a video to show the details of the spin configuration from all possible directions in \url{https://github.com/Chenhui-Song/Spherical-XY-Model}.

\section{Phase Transition Temperature}

\begin{figure}[th]
\centering
\includegraphics[width=5.5in]{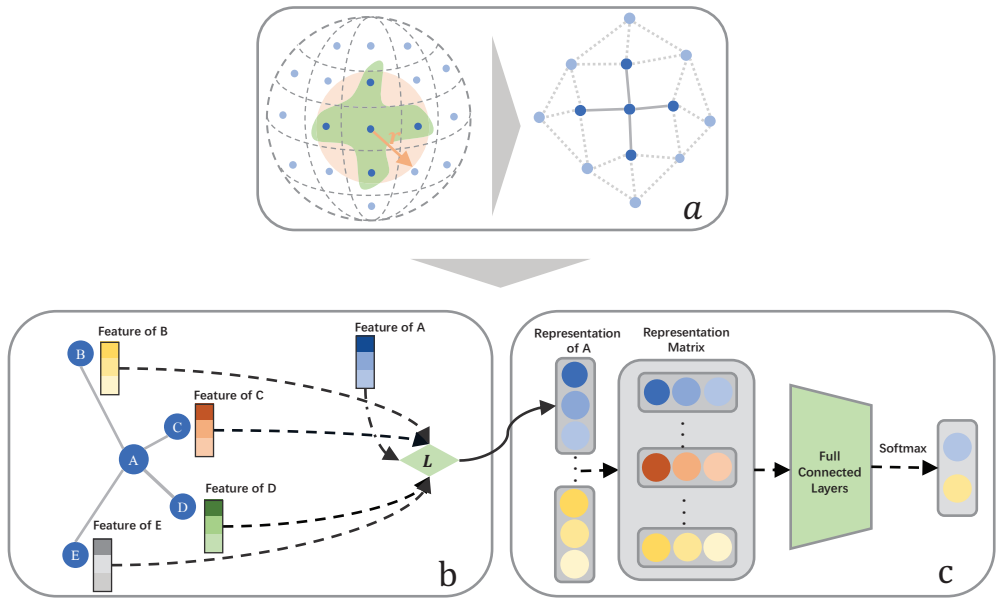}

\caption{Flow charts for phase classification of spherical XY model based on GCN. Panel a shows how to extract a graph structure from a spherical Fibonacci lattice. Panel b indicates that the input features are aggregated by the Laplacian matrix. Panel c shows that the classification confidences are finally obtained by using the fully connected layer.}
\label{Fig4c}
\end{figure}

To determine the phase transition temperature, a powerful tool is the newly developed paradigm, the machine learning method. There have been plenty of successful examples, among which the neutral network has particular advantages due to its strong generalization ability. Even the simplest three-layer fully connected neutral network (FCN) can well recognize different phases of Ising model\cite{MLNP17}. However, for topological phase transitions, FCN does not work equally well since it can not effectively capture the local spatial information, which may be directly related to the local orderedness. An alternative tool is the convolutional neutral network (CNN), which has proved its powerfulness in many situations\cite{CNNPRB18,CNNPRE19}. CNN relies on the convolutional kernels to extract local features from an input object, which usually requires the information distribution has a spatial translation invariance. Obviously, this is not satisfied by a spherical lattice. Even the Fibonacci lattice is just approximately uniform. Here we propose a new method for phase recognition based on the graph convolutional network (GCN)\cite{GCN20,GCN20b}.


GCN is a generalization to the traditional CNN based on the graph data structure. It can effectively aggregate the local spatial information, and thus
realize the classification of vertices and graphs. Details of GCN can be found in Appendix.\ref{app}, and Figure.\ref{Fig4c} shows the flow charts to classify different phases of the spherical XY model by GCN. Here we give a very simple introduction of the procedure. For a $N=1000$ spherical Fibonacci lattice, we set $r_\text{c}=1.1395$ such that most sites have 4 neighbors. We then map the lattice to a graph $\mathcal{G}$, and all information of $\mathcal{G}$ is stored in the degree matrix $\boldsymbol{D}$ and adjacent matrix $\mathcal{A}$, as shown in Fig.\ref{Fig4c}.a. The convolution is performed by using the Laplacian matrix $\boldsymbol{L}=\boldsymbol{D}-\mathcal{A}$. The input features are specified by the feature matrix $\boldsymbol{X}=\left(\boldsymbol{s}_1^T,\cdots,\boldsymbol{s}_N^T\right)^T\in\mathbb{R}^{N\times 3}$, where
$\boldsymbol{s}_i=(x_i,y_i,z_i)$ is the spin at site $i$. Next, we apply the random walk normalized Laplacian $\boldsymbol{L}^\text{rm}=\boldsymbol{D}^{-1}\boldsymbol{L}$ to aggregate the features, and use the ReLu function as an activation function to get the feature representation $\boldsymbol{H}\in\mathbb{R}^{N\times 1}$ (see Fig.\ref{Fig4c}.b)
\begin{align}
    \boldsymbol{H}=\text{ReLu}\left(\boldsymbol{L}^\text{rm}\boldsymbol{X}\boldsymbol{W}_h+\boldsymbol{b}_h\right)
\end{align}
where the weight $\boldsymbol{W}_h\in\mathbb{R}^{3\times 1}$ and the bias $\boldsymbol{b}_h\in \mathbb{R}^{N\times 1}$ are both learnable parameters. Finally, by using the fully connected layer and the $softmax$ function to aggregate $\boldsymbol{H}$, we get the output feature (see Fig.\ref{Fig4c}.c)
\begin{align}
    \hat{\boldsymbol{Y}}=softmax\left(\boldsymbol{H}^T\boldsymbol{W}_p+\boldsymbol{b}_p\right)
\end{align}
where $\boldsymbol{W}_p\in \mathbb{R}^{N \times 2}$ and $\boldsymbol{b}_p\in \mathbb{R}^{2}$ are also learnable parameters. The two components of the output feature $\hat{\boldsymbol{Y}}\in \mathbb{R}^2$ respectively correspond to the classification confidences of ordered phase ($p_\text{o}$) and disordered phase ($p_\text{d}$). It is reasonable to address $p_\text{o}=p_\text{d}$ at the critical temperature $T_\text{c}$.

\begin{figure}[th]
\centering
\includegraphics[width=3.4in]{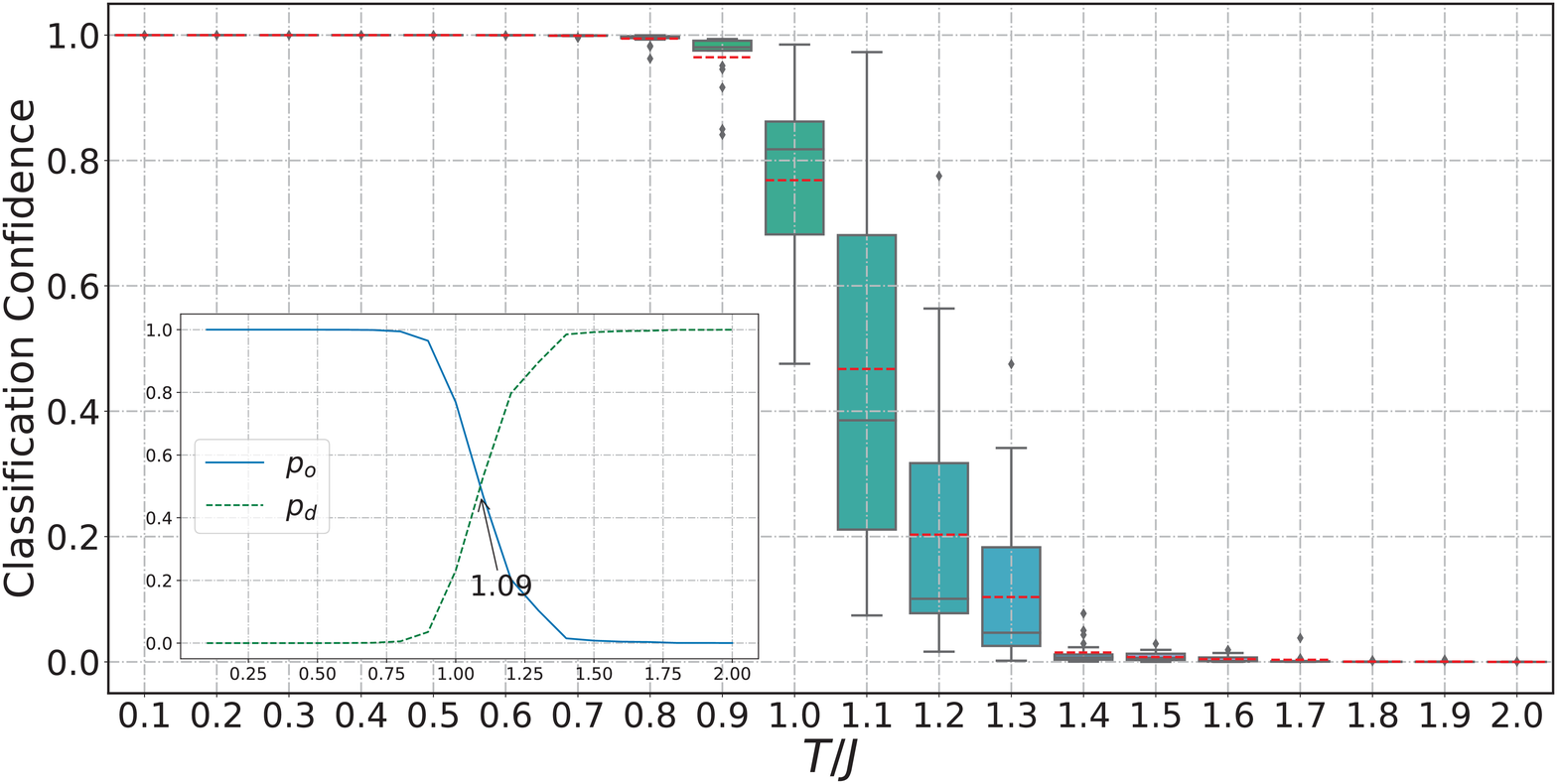}
\includegraphics[width=3.4in]{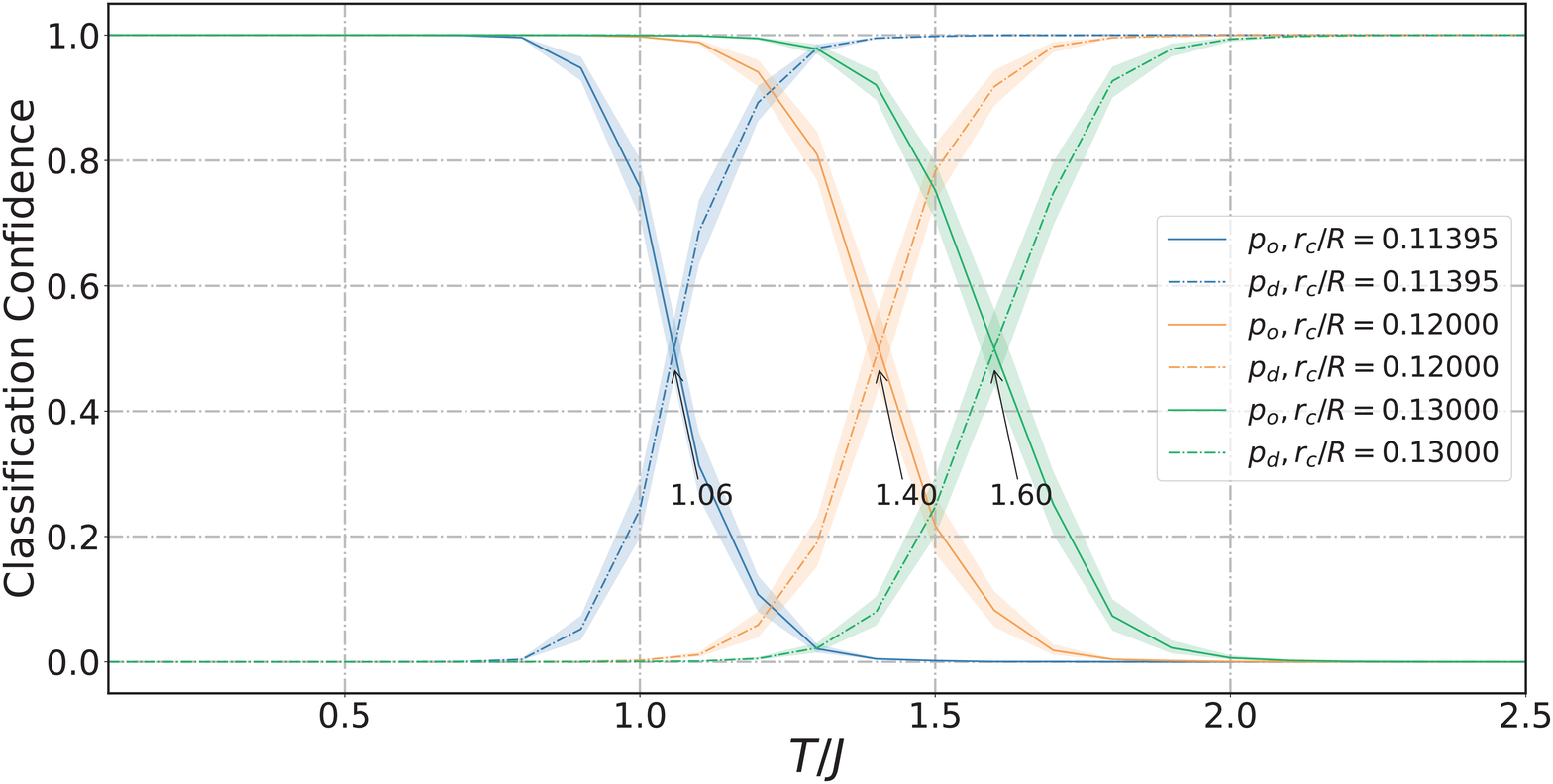}
\caption{(Left panel) Application of the GCN to the XY model on a $32\times 32$ square lattice, where $p_\text{o}$ is plotted vs temperature. Here the lower and top edges of each rectangle respectively label the lower and higher quartiles of calculated $p_\text{o}$, and the black-solid and red-dashed lines denote the corresponding median and average values. The inset shows how to determine the critical temperature. (Right panel) Critical temperature of a spherical Fibonacci-lattice XY model. The blue dashed, brown dot-dashed and green solid curves correspond to $r_\text{c}/R=0.11395$, 0.12 and 0.13 respectively. $p_\text{o}$ and $p_\text{d}$ are obtained by their average values.}
\label{Fig5b}
\end{figure}

Using MC simulations as the test sets of GCN, we can identify phases and phase transitions. We outline the numerical results in Figure \ref{Fig5b}.
We first test our GCN model by using the XY model on a $32\times32$ square lattice. We start with an arbitrary spin configuration and use MC simulations to give a spin sample, then we use the sample as the test set for the GCN and find an estimation of the classification confidences $p_\text{o}$ and $p_\text{d}$.
To obtain low temperature data samples, we perform MC simulations over the
temperature range $[0.01,0.05]J$ with an increment of $0.01J$ and repeat the simulations for 46 times. For high temperature data samples, we perform MC simulations over the
temperature range $[26.0,30.0]J$ with an increment of $1.0J$ and also repeat the simulations for 46 times. We then divide each of the 46 sampling sets into the training sets of 40 each, validation sets of 5 each, and test sets of 1 each. After the GCN model is trained with those samples, we apply it to the temperature range $[0.1,2.0]J$ with an increment of $0.1J$ and repeat the procedure 500 times at each temperature to give 100 sets of raw data.
In the left panel of Figure \ref{Fig5b}, the lower and upper edges of each rectangle respectively label the lower and higher quartiles of the data set, and the black-solid and red-dashed lines denote the corresponding median and average values (We only show the estimation of $p_\text{o}$).
We use the average values for the estimated $p_\text{o}$ and $p_\text{d}$. The critical temperature is determined by $p_\text{o}=p_\text{d}$.
The estimation of $T_c=1.09 J$ is very close to the result ($T_c=1.08J$) by the renormalization-group calculation~\cite{RGsquareXY93}. There are also very few statistically abnormal data labelled by the black dots, which basically have no influence on the final result.
Note here we consider a finite-size system. Strictly speaking, $T_c$ is not the Kosterlitz-Thouless transition temperature $T_\text{KT}$ in the thermodynamic limit, but the effective transition temperature at which the correlation length $\xi$ is comparable to the system size ($L=32$)~\cite{RGsquareXY93}. $T_c$ can be inferred from $T_\text{KT}$ via $T_c\approx T_\text{KT}+\frac{\pi^2}{c(\ln L)^2}$ where $c$ is a constant~\cite{RGsquareXY93}.

In the right panel, we apply the GCN to the XY model on a spherical Fibonacci lattice. Similarly, we repeat the procedure of the GCN for 100 times and then average those raw data.
We plot the classification confidences vs. temperature for $r_\text{c}/R=0.11395$, 0.12 and 0.13, and the corresponding critical temperatures are $T_\text{c}/J=1.06$, $1.40$, and $1.60$, respectively.
The increase of $T_\text{c}$ with $r_\text{c}$ is also reasonable since a larger $r_\text{c}$ means more neighbors are involved in the interactions. Therefore, more energetic thermal fluctuations are needed to unbind the vortices, signifying higher $T_\text{c}$.
We emphasize that similar inferences also hold for XY models on a planar lattice. For XY models on a large square-lattice ($L\rightarrow\infty$), each site has 4 neighbors and $T_\text{KT}/J\approx 0.898$~\cite{2DXYPSPRL88}. While for XY models on a large triangular-lattice, each site have 6 neighbors and $T_\text{KT}/J\approx 2.93$~\cite{TXYPRB94}. (Ref.~\cite{TXYPRB94} actually found $\beta_c\approx 0.683$ with $J=\frac{1}{2}$, implying $T_\text{KT}/J\approx 2.93$.) Hence, they demonstrate the same influence of the number of neighbors on the transition temperature.
The numerical results indicate the SGD optimizer works very well while others, like the Adam, do not produce accurate results.

\section{Vortex dynamics}
\begin{figure}[th]
\centering
\includegraphics[width=3.4in]{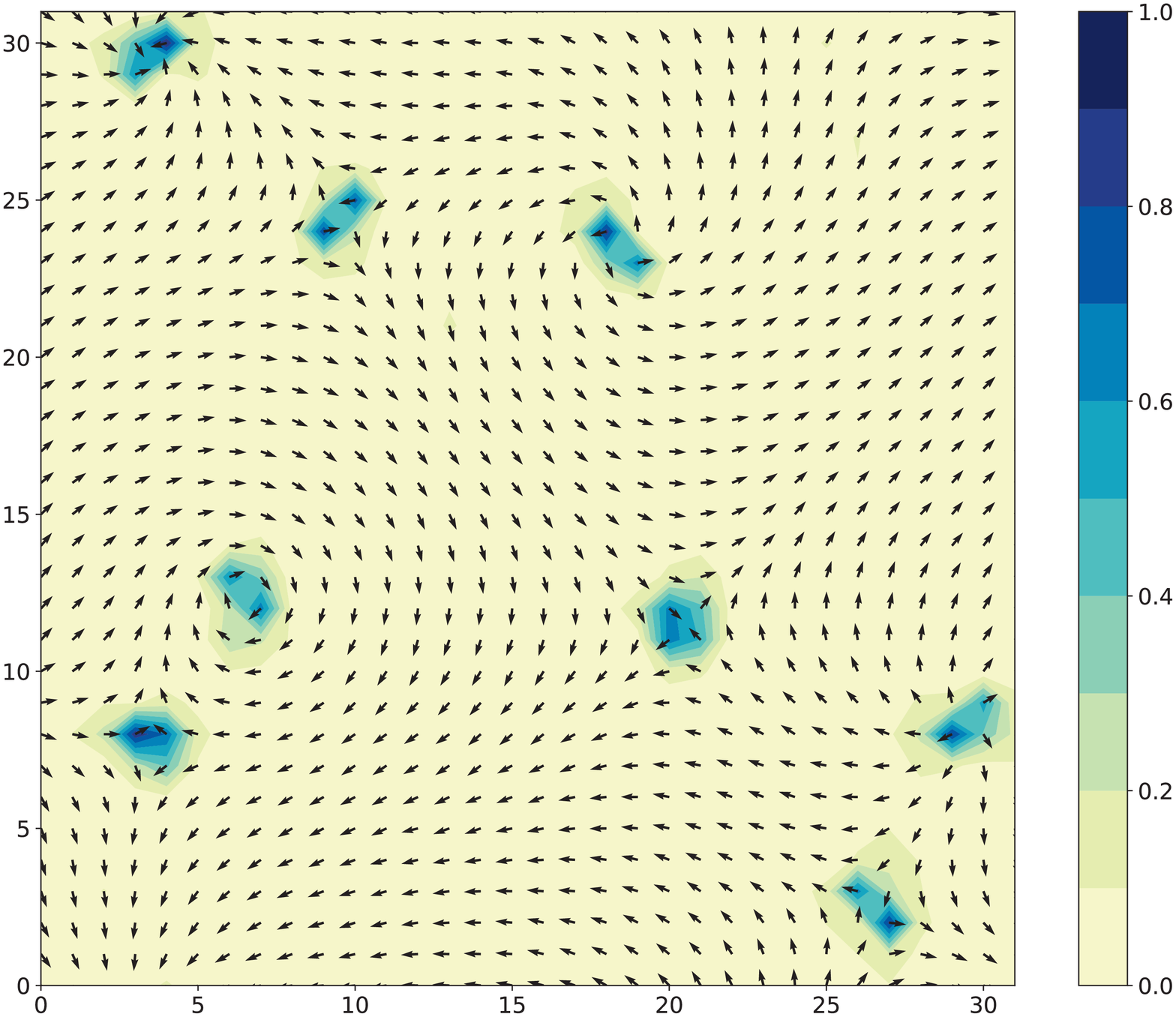}
\caption{The local disorder of the vortices in the XY model on a 2D square lattice, where $r_c/a=1.0$, and $T/J=0.005$.}
\label{Fig5}
\end{figure}

We also investigate the motion of the vortices on the sphere during a MC annealing. In fact, the number of MC attempts is related to the real time via a certain function $f$: $n=f(t)$\cite{XYD83}. Hence the MC simulation, which brings a system to equilibrium from a nonequilibrium starting point, is equivalent to the spontaneous relaxation dynamics governed by the Langevin equation\cite{XYD83,XYPRB87}. Physically, the trajectory actually describes the time-dependent evolution of vortices on a sphere, and $n=f(t)$ can be thought of as a reparametrization of the evolutionary path. To trace out the motion of vortices, we need to locate where a vortex is at first. Previous discussions have shown that the GCN has a superior performance on recognizing ordered/disordered phases, even with only one convolutional layer. This has a qualitative explanation. For a one-convolutional-layer GCN, the Laplacian matrix realizes the feature aggregation of the neighbors for each site, i.e.
\begin{align}
    \boldsymbol{R}=\boldsymbol{L}^\text{rm}\boldsymbol{X}=\left(\boldsymbol{r}_1^T,\cdots,\boldsymbol{r}_N^T\right)^T
\end{align}
where $\boldsymbol{r}_i=\frac{1}{\left\|\mathcal{E}_i\right\|}\sum_{j \in \mathcal{E}_i }\left(\boldsymbol{s}_j-\boldsymbol{s}_i\right)$, and $\mathcal{E}_i$ is the set of neighbors of site $i$. This provides a good indicator to probe the local orderedness. At sites where there are no vortices, the spin changes slowly, thus $\left\|\boldsymbol{r}_i\right\|$ is close to zero. On the contrary, at sites where there exist vortices, the spin changes dramatically, and $\left\|\boldsymbol{r}_i\right\|$ must be finite. Hence, we define $\xi_i=\left\|\boldsymbol{r}_i\right\|$ as the the local disorderedness at site $i$. To detect vortices, we set a threshold $\Delta=\gamma\left\|\bar{\boldsymbol{r}}\right\|$, where $\left\|\bar{\boldsymbol{r}}\right\|=\frac{1}{N}\sum_i\left\|\boldsymbol{r}_i\right\|$ is the average local disorderedness, and $\gamma$ is an adjustable parameter. As a check of this method, we apply it to a $30\times 30$ square-lattice XY model and present the result in Fig.\ref{Fig5}. Evidently, all vortices are precisely probed by their local disorderedness.

\begin{figure}[th]
\centering
\includegraphics[width=1.9in]{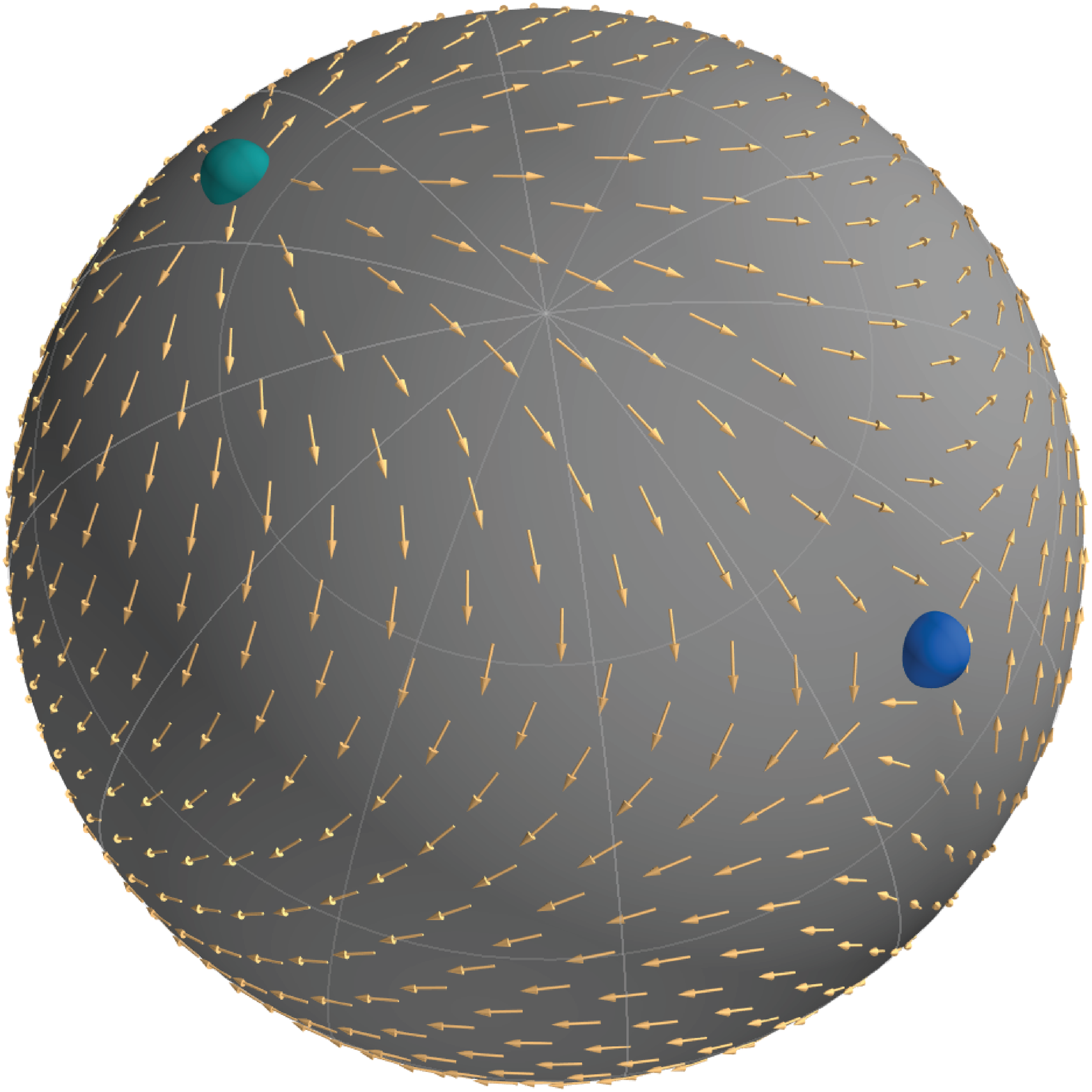}
\includegraphics[width=1.9in]{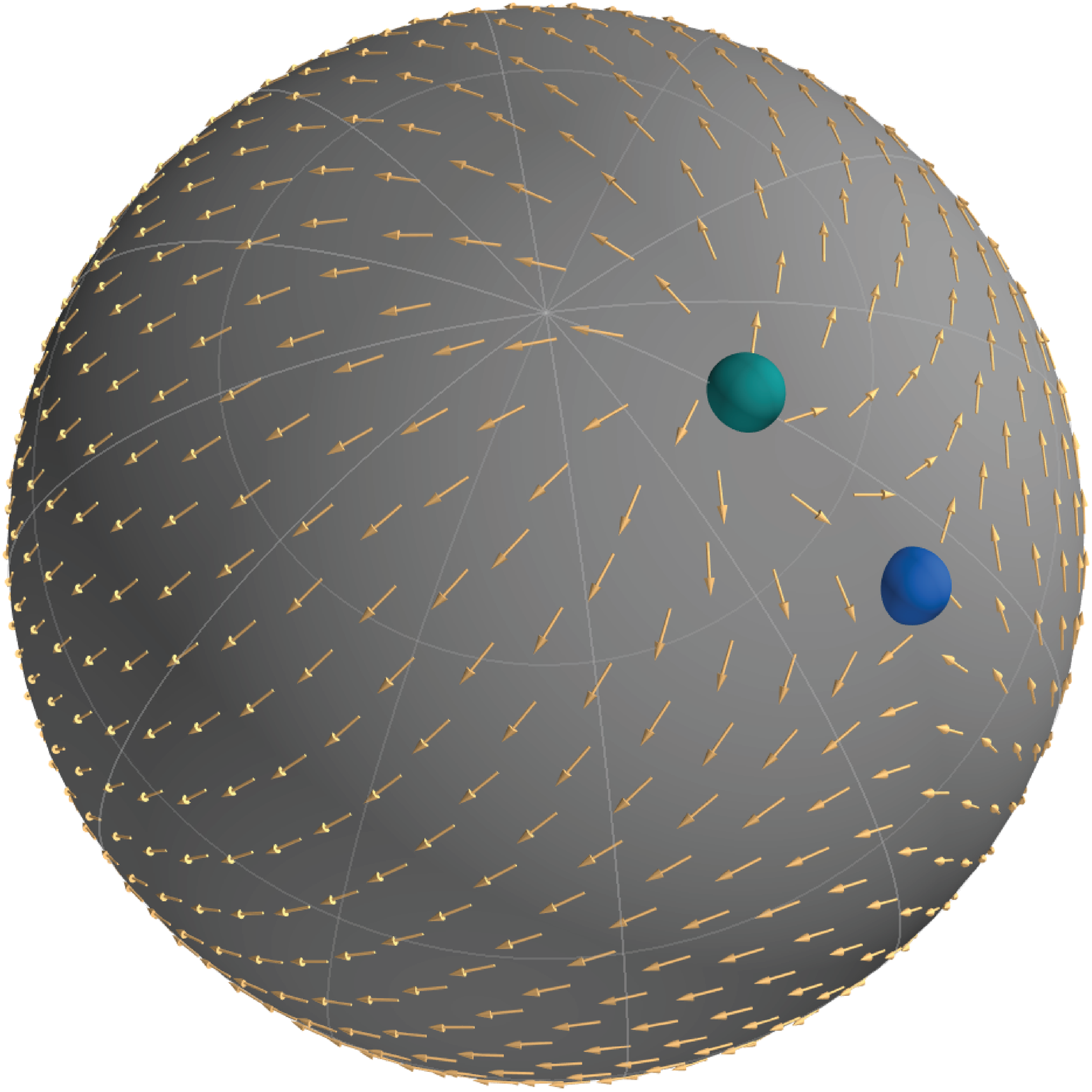}
\includegraphics[width=1.9in]{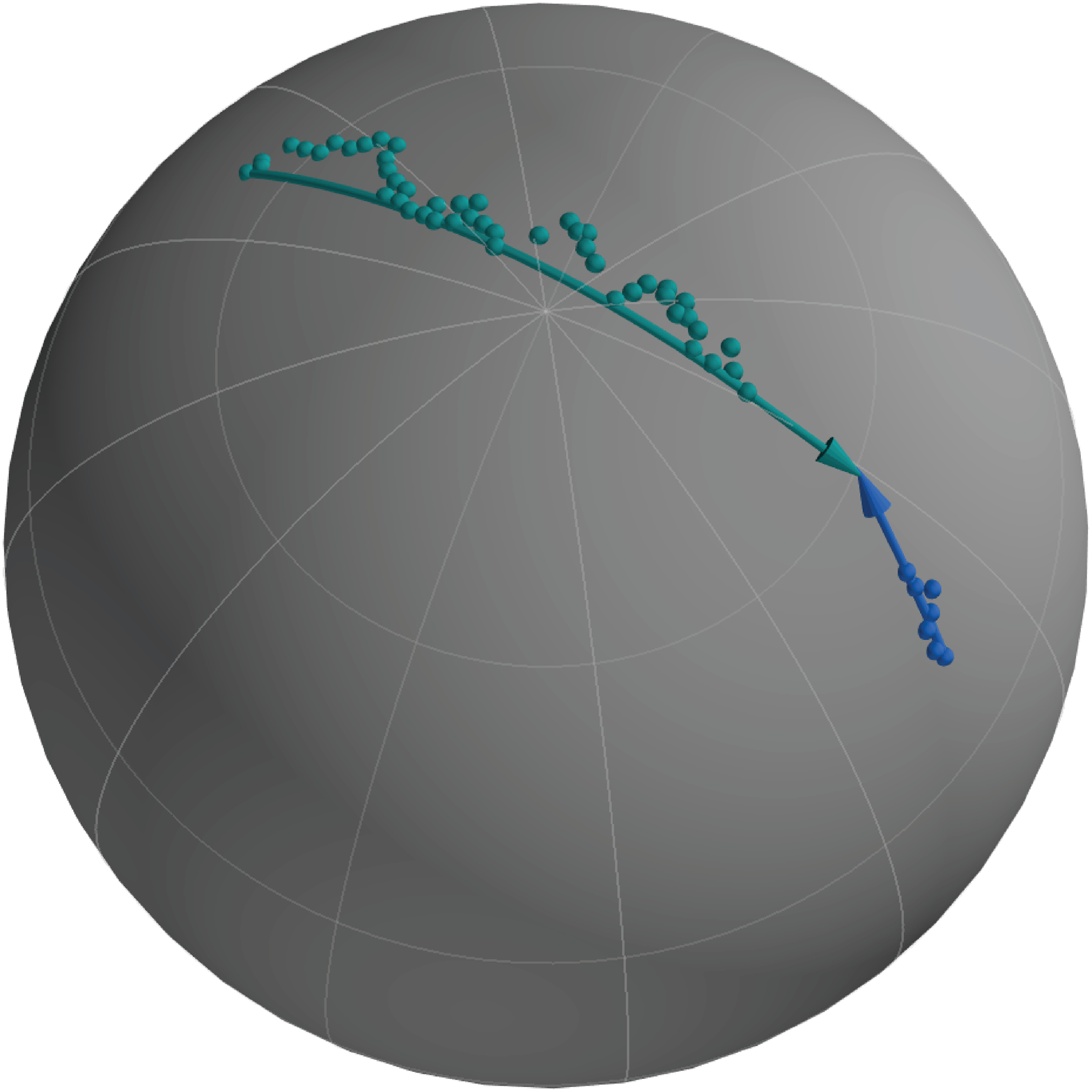}\\
\includegraphics[width=5.7in]{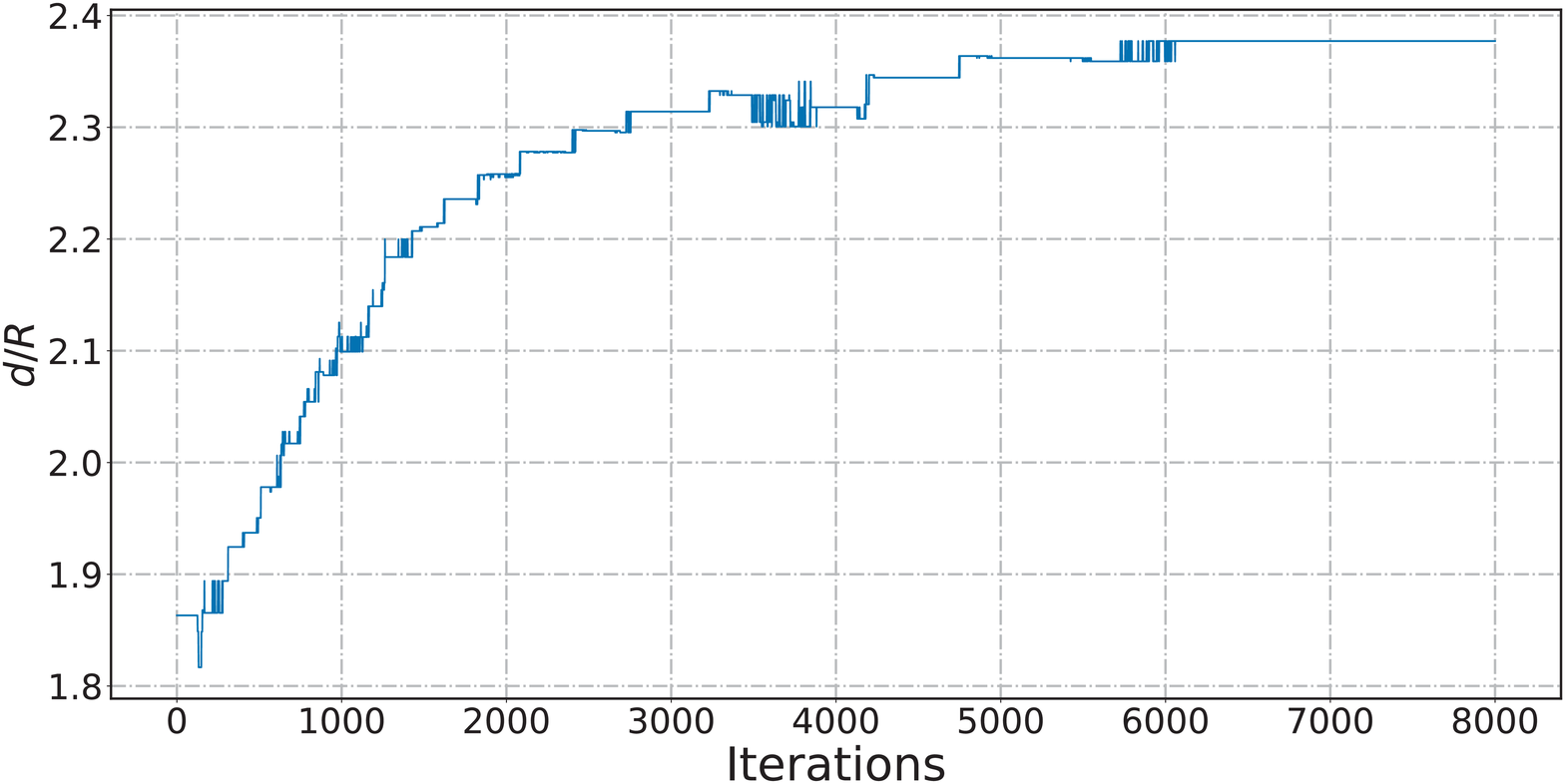}
\caption{(Top panels) Snapshots of two annihilating vorices (left and middle) and their trajectories (right). (Lower panel) Distance between two vortices as a function of Monte Carlo steps. Here one ``iteration'' is equivalent to 10000 steps, as stated in the main text.}
\label{Fig6}
\end{figure}

We then apply the method to the spherical XY model. After finding the vortices, their evolutionary paths can further be traced out by a data clustering algorithm.
Specifically, we first filter out the coordinates such that $\xi_i>\Delta$ at every 10000 steps, which obviously belong to certain vortices. We label the set of these coordinates by $P$, and further use a well known machine learning method, the density-based spatial clustering of applications with noise
(DBSCAN)\cite{DBSCAN}, to group these coordinates by vortices:
\begin{align}
    \left\{P_1,\cdots,P_k\right\}=\text{DBSCAN}(\text{Minpts},r,P).
\end{align}
Here Minpts and $r$ are two required parameters, $k$ is the number of different vortices, and $P_i(i\in \left\{1,\cdots,k\right\})$ is the classified coordinate-set, each of which belongs to a single vortex. The position of a vortex is given by
\begin{align}
    \boldsymbol{x}_i^v=\mathop{argmin}\limits_{\left\|\boldsymbol{x}\right\|=R}\sum_{\boldsymbol{x}^p\in P_i}\left\|\boldsymbol{x}-\boldsymbol{x}^p\right\|^2.
\end{align}
With these tools, we finally obtain the dynamics of the vortices on a spherical surface. We visualize our results with $N=1000$, $r_\text{c}/R=0.15$, and $T/J=5.0\times 10^{-4}$ in Fig. \ref{Fig6}.
Previous results have shown that initially there are multiple vortices but eventually only two of them survive, as others merge during the MC annealing. This is also true here. Figure. \ref{Fig6} show the merging process of a pair of vortices.
In the upper row, we present the merging process of a pair of vortices with different topological charges.
The left plot shows an early vortex configuration, in which the two vortices are far from each other, and the middle plot shows an intermediate state,
in which the two vortices move close to each other. The right one shows the merging path of the two vortices.
The lower panel shows the distance between the two vortices, which finally survive, as a function of the MC step. Here the distance is evaluated along the great circle (geodesic curve) connecting the two vortices on the sphere.

\section{Conclusion}

In this paper, we have presented distinct features of the XY model on a spherical surface, inspired by cold-atom experiments in microgravity
environments, such as the International Space Station. Using the Fibonacci lattice, an almost uniform lattice on a sphere, along with MC simulations and machine learning techniques like the GCN model, our results predict the vortex distributions and phase transition temperatures. Moreover, the DBSCAN clustering method allows us to visualize the evolution paths of the vortices during a MC annealing. The methods and analyses shed light on future experiments of engineered systems with spherical geometries.

H. G. was supported by the National Natural Science Foundation
of China (Grant No. 12074064).

\appendix
\section{Introduction to GCN}\label{app}
\subsection{Problem Formulation}

Our aim is to distinguish the ordered phase at low temperatures and disordered phase at high temperatures of the spherical XY model by using GCN. Specifically, we are going to obtain a classification function $\hat{\boldsymbol{Y}}=F(\boldsymbol{X};\Theta)$ by data training, which maps the feature matrix $\boldsymbol{X}$ containing all spin information into the label space $\hat{\boldsymbol{Y}}=(p_\text{o},p_\text{d})$. Here $p_\text{o}$ and $p_\text{d}$ respectively give the classification confidences of the ordered and disordered phases. The training process is to adjust the set $\Theta$ of parameters such that the classification result $\hat{\boldsymbol{Y}}$ is as close to the real label $\boldsymbol{Y}$ as possible.

\subsection{Preliminaries}
Graph is a powerful data structure for processing relational information, which can effectively realize the clustering of neighbor information. A graph  $\mathcal{G}$ comprises a set $\mathcal{V}$ of vertices and a set $\mathcal{E}$ of edges. For two vertices $v_i$, $v_j\in \mathcal{V}$, if there exists an edge $e_{ij} \in \mathcal{E}$, $v_i$ and $v_j$ are said to be connected by $e_{ij}$ in the graph drawing in $\mathcal{G}$.
The number of vertices to which a vertex $v_i$ is linked is defined as the degree $d_j$ of $v_i$.
Explicitly, $\mathcal{G}$ can be represented by the adjacent matrix $\mathcal{A}$, of which the entries are given by $\mathcal{A}_{ij}=1$ if $e_{ij}\in \mathcal{E}$, or 0 otherwise. The feature of each vertex $v_i$ can be expressed by a $d$-dimensional vector $\mathbf{x}_i$. Thus, for all vertices, their features can be represented by a matrix $\boldsymbol{X}\in \mathbb{R}^{n\times d}$, where $n$ is the degree of $\mathcal{G}$.

GCN is a generalization of the traditional CNN, which can be used to classify vertices and graphs. There are many ways to realize GCN, which may be mainly classified into spatial-based and spectral-based convolutions. Both of them can be realized with the help of the Laplacian matrix $L$, defined by $\boldsymbol{L}=\boldsymbol{D}-\mathcal{A}$, where $\mathcal{A}$ is aforementioned adjacent matrix, and $\boldsymbol{D}$ is a diagonal matrix called degree matrix, of which the diagonal elements are the degrees of vertices. The definition of Laplacian matrix is not unique, and the frequently-used definitions include the symmetric normalized Laplacian $\boldsymbol{L}^\text{sym}=\boldsymbol{D}^{-\frac{1}{2}}\boldsymbol{L}\boldsymbol{D}^{-\frac{1}{2}}$ and the random walk normalized Laplacian $\boldsymbol{L}^\text{rm}=\boldsymbol{D}^{-1}\boldsymbol{L}$.

The spatial-based GCN can be expressed by
\begin{align}
    \boldsymbol{H}^{l+1}=\sigma\left(\boldsymbol{LH}^l\boldsymbol{W^l}\right)
\end{align}
where $\boldsymbol{W}$ is the set of learnable parameters, $\boldsymbol{H}^l$ is the feature of the $l$-th layer (Note $\boldsymbol{H}^0$ represents the input feature), and $\sigma(\cdot)$ is the activation function. In this method, the feature of each vertex is obtained by aggregating the information of its neighbours.

The spectral-based GCN is realized by taking the spectral decomposition of the Laplacian matrix (i.e. the Fourier transformation) firstly and performing the associated inverse transformation secondly. It can be expressed by
\begin{align}
    \boldsymbol{H}^{l+1}=\sigma(\boldsymbol{U}g_\theta(\boldsymbol{\Lambda})\boldsymbol{U}^T\boldsymbol{H}^l)
\end{align}
where $\boldsymbol{U}$ is the matrix comprised by the eigenvectors of $\boldsymbol{L}$, $\boldsymbol{\Lambda}$ is the diagonal matrix comprised by the eigenvalues of $\boldsymbol{L}$, and the convolutional kernel $g_\theta(\boldsymbol{\Lambda})$ represents the set of learnable parameters. A widely-adopted choice of $g_\theta$ is the Chebyshev polynomial, which can reduce the number of steps of eigendecomposition and is thus more suitable for large-scale graph networks.

\subsection{Methodology of Phase Classification}

Here we illustrate how to apply GCN to recognize different phases of the spherical XY model, which is shown schematically by flow charts in Fig.\ref{Fig4c}.
The convolutional kernel of traditional CNNs can capture the local spatial information, which corresponds to the short-range orderedness of square XY models at low temperatures. This is why traditional CNN works well in these situations. Note the key point in phase-recognition is to effectively aggregate
the local spatial information. For XY models, the first step toward this is to establish the neighbour relation of each spin, which is guaranteed by the homogeneity of the square lattice. However, no such lattices exist on a spherical surface, and the Fibonacci lattice is only an approximate uniform lattice which has no translation invariance. Thus, the traditional CNN does not apply here. Fortunately, GCN instead provide an ideal structure to characterize the neighbour relation of irregular lattices. For a spherical Fibonacci lattice, all sites naturally comprise a set of vertices $\mathcal{V}$. Moreover, by choosing a critical radius $r_\text{c}$ such that $e_{ij}\in \mathcal{E}$ if $||\mathbf{r}_i-\mathbf{r}_j||<r_\text{c}$, a graph $\mathcal{G}$ can be constructed. To better capture local features, the value of $r_\text{c}$ must be carefully selected such that each site is only linked to nearby sites. For example, when $N=1000$, the (almost) best choice is $r_\text{c}=0.11395R$, shown in Fig.\ref{Fig4c}.a.
In such a spherical XY model, the spin at site $i$ is denoted by a unit vector $\boldsymbol{s}_i=(x_i,y_i,z_i)$ with $i=1,2,\cdots, N$. Thus, the feature matrix is given by $\boldsymbol{X}=\left(\boldsymbol{s}_1^T,\cdots,\boldsymbol{s}_N^T\right)^T\in\mathbb{R}^{N\times 3}$. In this paper, we use the random walk normalized Laplacian $\boldsymbol{L}^\text{rm}=\boldsymbol{D}^{-1}\boldsymbol{L}$ to aggregate the features, and apply the ReLu function as the activation function, as shown in Fig.\ref{Fig4c}.b
\begin{align}
    \boldsymbol{H}=\text{ReLu}\left(\boldsymbol{L}^{rm}\boldsymbol{X}\boldsymbol{W}_h+\boldsymbol{b}_h\right)
\end{align}
where $\boldsymbol{W}_h\in\mathbb{R}^{3\times 1}$ and $\boldsymbol{b}_h\in \mathbb{R}^{N\times 1}$ are both learnable parameters and $\boldsymbol{H}\in\mathbb{R}^{N\times 1}$ is the feature representation. We further use the full connected layer and the $softmax$ function to aggregate the representation $\boldsymbol{H}$, as demonstrated in Fig.\ref{Fig4c}.c, which can also be expressed by
\begin{align}
    \hat{\boldsymbol{Y}}=softmax\left(\boldsymbol{H}^T\boldsymbol{W}_p+\boldsymbol{b}_p\right)
\end{align}
where $\boldsymbol{W}_p\in \mathbb{R}^{N \times 2}$ and $\boldsymbol{b}_p\in \mathbb{R}^{2}$ are both learnable parameters, and the output $\hat{\boldsymbol{Y}}\in \mathbb{R}^2$ give the classification confidences that the input features respectively correspond to ordered/disordered phases.

\subsection{Training Procedure}

1. \textit{General settings}: The trial data come from MC simulations. As an example, we generate a $N=1000$ Fibonacci lattice on a spherical surface of radius $R$. On the tangent plain of each lattice site, a randomly-oriented spin is assigned initially. When mapping the lattice into a graph, we choose $r_\text{c}/R=0.11395$ and use the cross entropy function as the loss function.

2. \textit{Data settings}: At first, we give a very rough estimation of the critical temperature $T_\text{c}$ by several runs of MC simulations, which shows $T_\text{c}/J\sim 2.0$. Then, at temperatures far below $T_\text{c}$, we perform MC simulations over the temperature range $[0.005, 0.05]J$ with an increment of $0.005J$ and get a data set of the spin distributions in the low-temperature ordered phase. We repeat the simulations over the same temperature range for 26 times and get 26 sampling sets.
Similarly, at temperature far above $T_\text{c}$, we perform MC simulations over the temperature range $[25.5, 30]J$ with an increment of $0.5J$ and get a data set of the spin distributions in the high-temperature disordered phase. We also repeat the simulations to get 26 sampling sets. Finally, we split them into the training set, validation set, and test set, which respectively have 20, 5 and 1 sampling set for the ordered and disordered phases.

3. \textit{Critical temperature}: A prediction of the critical temperature can be achieved if the GCN model is trained sufficiently. We sample the data in a temperature range, in which $T_\text{c}$ is included, and input those features into the GCN model. The output gives the confidences that the input features correspond to the ordered ($p_\text{o}$) and disordered ($p_\text{d}$) phases. Obviously, $p_\text{o}+p_\text{d}=1$. We extract the critical temperature as the location when $p_\text{o}=p_\text{d}$. In order to suppress the errors, we carry out multiple simulations and classifications with the same set of parameters and average the corresponding classification confidences. For example, we execute our algorithm in the range $[0.1, 2.5]J$ with an increment of $0.1J$ and repeat the procedure 100 times to average the confidences for $r_\text{c}/R=0.11395$, 0.12, and 0.13, respectively.

4. \textit{Numerical Results}: When testing the GCN model with the samples from the test sets, it gives the correct classification results of $\ge 99.9\%$ classification confidences and 100$\%$ accuracy, even with only one convolutional layer.

\bibliographystyle{apsrev}
\bibliography{Review,Review1}
\end{document}